\documentclass[10pt,a4paper,onecolumn,aps,nofootinbib,showpacs,floatfix]{revtex4}

\usepackage[T1]{fontenc}
\usepackage[cp1250]{inputenc}
\usepackage{graphicx}
\usepackage{placeins}
\usepackage{dsfont}
\usepackage{amsmath,amssymb,graphics}
\usepackage{color}
\usepackage{bm}

\begin{document}

\title{Nucleation, condensation and $\lambda $-transition within the complex network: An application to real-life market evolution}
\author{M. Wili{\'n}ski}
\email{mateusz.wilinski@fuw.edu.pl}
\author{B. Szewczak}
\email{b.szewczak@gmail.com}
\author{T. Gubiec}
\email{Tomasz.Gubiec@fuw.edu.pl}
\author{R. Kutner,}
\email{(correspondence)  erka@fuw.edu.pl}
\affiliation{Institute of Experimental Physics Faculty of Physics, University of Warsaw  \\ Ho\.za 69, PL-00681 Warsaw, Poland}
\author{Z.R. Struzik}
\email{zbigniew.struzik@p.u-tokyo.ac.jp}
\affiliation{The University of Tokyo, 7-3-1 Hongo, Bunkyo-ku, Tokyo 113-0033, Japan \\ and \\
RIKEN Brain Science Institute, 2-1 Hirosawa, Wako-shi 351-0198, Japan}

\begin{abstract}
We fill a void in merging empirical and phenomenological characterisation of the dynamical phase transitions in complex systems by identifying three of them on real-life financial markets. We extract and interpret the empirical, numerical, and semi-analytical evidences for the existence of these phase transitions, by considering the Frankfurt Stock Exchange (FSE), as a typical example of a financial market of a medium size. Using the canonical object for the graph theory, i.e. the Minimal Spanning Tree (MST) network, we observe: (i) The initial phase transition from the equilibrium to non-equilibrium MST network in its nucleation phase, occurring at some critical time. Coalescence of edges on the FSE's transient leader is observed within the nucleation and is approximately characterized by the Lifsthiz-Slyozov growth exponent; (ii) The nucleation accelerates and transforms to the condensation process, in the second phase transition, forming a logarithmically diverging $\lambda $-peak of short-range order parameters at the subsequent critical time; (iii) In the third phase transition, the peak logarithmically decreases over three quarters of the year, resulting in a few loosely connected sub-graphs. The 
$\lambda $-peak, resembling the continuous phase transition from the normal fluid I $ ^4$He to the superfluid II $ ^4$He, is reminiscent of a 
non-equilibrium superstar-like superhub or a `dragon-king' effect, abruptly accelerating the evolution of the leader company. The complexity of the MST
is reduced then to the nonlinearity present in deterministic coarse-grain ('macroscopic') dragon-king dynamic equation derived in tha paper.
All these phase transitions are supported by the few richest vertices, which drift towards the leader and provide the most of the edges increasing the leader's degree. Thus, we capture an amazing phenomenon, likely of a more universal character, where a peripheral vertex becomes the one which is over dominating the complex network during an exceptionally long period of time.
\end{abstract}

\pacs{89.65.Gh, 89.20.-a, 89.65.-s, 05.40.-a, 02.30.Mv, 02.50.-r, 02.50.Ey, 02.50.Ga}

\maketitle

\section{Introduction}\label{section:hypothesis}

For one-and-half decade, physicists have been intensively studying structural and topological properties of complex networks 
\cite{RAALB0,RAALB,BCLM0,DFPV,DGM,LBS,MFA0,KwDr,MWH,DiHe,ThB,RNMJK,DKORak,SH} (and refs. therein) in order to understand the mechanisms responsible for the evolution of real-world complex systems and their miscellaneous consequences. Arguably, one of the most exclusive among these, is the condensation phenomenon, together with a $\lambda $-transition (or temporal $\lambda $-peak) associated with this phenomenon -- which hitherto have never been observed in a real-world network \cite{SND}. By the term `temporal $\lambda $-peak' we understand here the temporal shape of a short-range order parameter (or other parameters, e.g. of higher order, such as network characteristics) resembling the Greek letter $\lambda $. This terminology is in analogy to the $\lambda $-peak of the heat capacity vs. temperature formed by the $\lambda $-transition between the normal I $ ^4$He and superfluid II $ ^4$He components. This analogy is developed further in the text. Here, we demonstrate the first evidence for such a real-life condensation phenomenon with an associated $\lambda $-transion, together with a preceeding phase of nucleation growth. Furthermore, we here investigate in empirical and phenomenological ways, a complete phase diagram for these intriguing dynamic phase transitions in real-world complex networks. 

Variety of complex network models show the phenomenon of condensation, where a finite fraction of structural elements in the network (edges, triangles, etc.) turn out to be aggregated into an ultra compact sub-graph (e.g. a star-like structure of edges), having size distinctly smaller than the size of the network \cite{DGM} (and refs. therein), yet sufficiently large strongly to dominate all other local structures present in this network. It is particularly convenient to measure this size using a mean occupation layer (MOL), introduced by Onnela-Chakraborti-Kaski-Kert\'esz \cite{OCKK1,OCKK2} for study of the 
S\&P 500 index in the vicinity of Black Monday (October 19, 1987) and also in the vicinity of the currency crisis in January 1, 1998. In this context, MOL can play the role of a temporal short-range order parameter\footnote{As we will see, a more appropriate name would be \emph{disorder parameter}, because the lesser is its value, the more star-like the local structure.} sufficiently sensitive to the local structure of a complex network \cite{WSGKS,SGKS}. 

In this work we consider empirical, undirected, canonical Minimal Spanning Tree (MST) being of a correlation based network of assets' returns, where multiple connections and loops are not allowed \cite{BeBo,RNM,BCLMVM,MS,BLM,VBT,KKK,BCLM,TMAM,TCLMM,ISI,IHSICh} and the number of vertices, $n$, and edges, $n-1$, are fixed. For such a network, the inter-node distance is defined as $d(i,j)\propto \sqrt{1-C(i,j)}$, for any pair of vertices. Other popular metrics were also used for comparison; however, the obtained results are practically undistinguishable within the resolution considered in our work. Although we verified in an empirical way, that during the MST evolution only the positive correlations participate in its construction, the transformation from the Pearson's correlation coefficient, $C(i,j)$, to distance, $d(i,j)$, was necessary, because the correlation coefficient does not obey the axioms of a metric (or even axioms of a subdominant ultrametric distance \cite{RNM}). This verification was made at each time step after the construction of a temporal MST network from a complete empirical graph. We had the opportunity to compare $n-1$ temporal MST correlation based distances with all $n(n-1)/2$ ones, which also contain several anti-correlation based distances. However, we can consider our temporal MSTs as compact ones as the covariance matrix is in our case only positive and semi-definite \cite{BCLM} (and refs. therein). Furthermore, we expect that other correlation based networks, e.g. such as the Threshold Networks and Hierarchical Networks, will give very similar results, if the threshold value and the  number of hierarchy levels are assumed to be sufficiently realistic quantities \cite{NMHL} (and refs. therein).

In spite of the algorithmic simplicity of the MST construction, the dynamics of MST network is still puzzling at microscales \cite{BCLM}, because relocation of edges (links) during the evolution of the MST network, potentially involves a rearrangement of the entire graph. This defines the nonseparability feature, which makes the Granger causality analysis inadequate for the identification of causation between variables represented as time-series \cite{SMY}. As such, this evolution is indeed a collective phenomena \cite{BCLM}, where only the algorithmic (and not an analytic) 
recipe\footnote{The most popular are both Prim and Kruskal algorithms \cite{SGKS} (and refs. therein) in this context, as no analytic routine is known.} defines the network's single time-step transformation. We develop (by neglecting fluctuations, cf. Sec. VIII.1 in Ref. \cite{NGvKamp}) such a phenomenological description, where the network rearrangement is derived from, and formally equivalent to the phenomenological `macroscopic' evolution equations, where the transition probabilities involved are verified in both semi-analytical and empirical ways. A `microscopic', qualitative explanation of the nucleation and condensation processes involved, were formulated through a detailed observation and analysis of the evolution of the MST network on the properly prepared sequence of snapshot frames (i.e. a `movie'), where active nodes and edges were suitably marked to make systematic tracing possible.

We demonstrate a diachronic \cite{BCK} approach to condensation, complementary to those considered by Albert-Barab\'asi \cite{RAALB} (and refs. therein) and Dorogovtsev-Goltsev-Mendes \cite{DGM} (and refs. therein). That is, we focus on the birth and on the death of condensation as dynamic phenomena, which occur as a result of the dynamic $\lambda $-transition, between two non-equilibrium states of a complex network. This leads to the condensate, which arises as a temporal superstar-like structure. This structure is manifested in a form of a temporal singularity\footnote{Obviously, this singularity is truncated because of finite size effects.} of the maximal vertex degree defining the dynamic $\lambda $-peak, where both its sides diverge logarithmically. At this time we believe, this is the first work confirming a birth and a death of a condensate, manifested through the temporal $\lambda $-peak, in a real-world complex network representing dynamic phenomena on a stock market. 

The condensation process is preceded by the nucleation growth entered by the MST network as a result of the continuous state transition from the equilibrium phase. As usual, the equilibrium state is defined by the detailed balance conditions (DBC) -- they are valid for the background of vertex degrees, further also referred to as the market `plankton'. The power law  distribution observed, of these plankton degrees -- not exceeding a dozen or so, is considered as the equilibrium distribution. For the initial, equilibrium state, all vertices obey this DBC criterion, while for the non-equilibrium states we trade with two kinds of `fluids' -- the equilibrium background (the basic fluid) and giant fluctuations of few vertex degrees (an `excited' fluid)
consisting of few remaining vertices exhibiting giant fluctuations of their degrees.   

We foresee that our results containing universal aspects, complementary in nature to those in Ref. \cite{KwDr} (and refs. therein), will provide a new impact to the modeling of dynamic structural and topological phase transitions and critical phenomena\footnote{From years 2005 to 2008 the Project of NEST Action in 6th EU Framework Programme entitled \emph{Critical Events in Evolving Networks} was developed. This Project was coordinated by Janusz Ho{\l}yst from Center of Excellence for Complex Systems Research and Faculty of Physics Warsaw University of Technology. The Project provided results from broad range of disciplines (e.g. econo- and sociophysics) by using complex evolving networks and phase transitions as generic tools (for details please refer to http://www.creen.org).} on financial markets \cite{DGM,DiSo,WH,PBX}.

The paper is organized as follows. In Sec. \ref{section:PhaMST} the main goal is defined together with a systematic presentation of our empirical results, constituting a basis for the phenomenological considerations following in the two subsequent sections \ref{section:keqpoor} and \ref{section:MeSZG}. That is, in Sec. 
\ref{section:keqpoor} the dynamics of poor vertices obeying a separability principle \cite{SMY} is studied, while Sec. \ref{section:MeSZG} concerns the complex critical dynamics of the richest vertex, as a reminiscence of network complexity. In Sec. \ref{section:Sumconcl} we discuss and summarize our results as well as highlight the most significant phenomenon found in the evolution of the MST network as a simple but sufficiently realistic and complex reference one.

\section{Phenomenology of the MST network evolution}\label{section:PhaMST}

\begin{figure}
\begin{center}
\bigskip
\includegraphics[width=160mm,angle=0,clip]{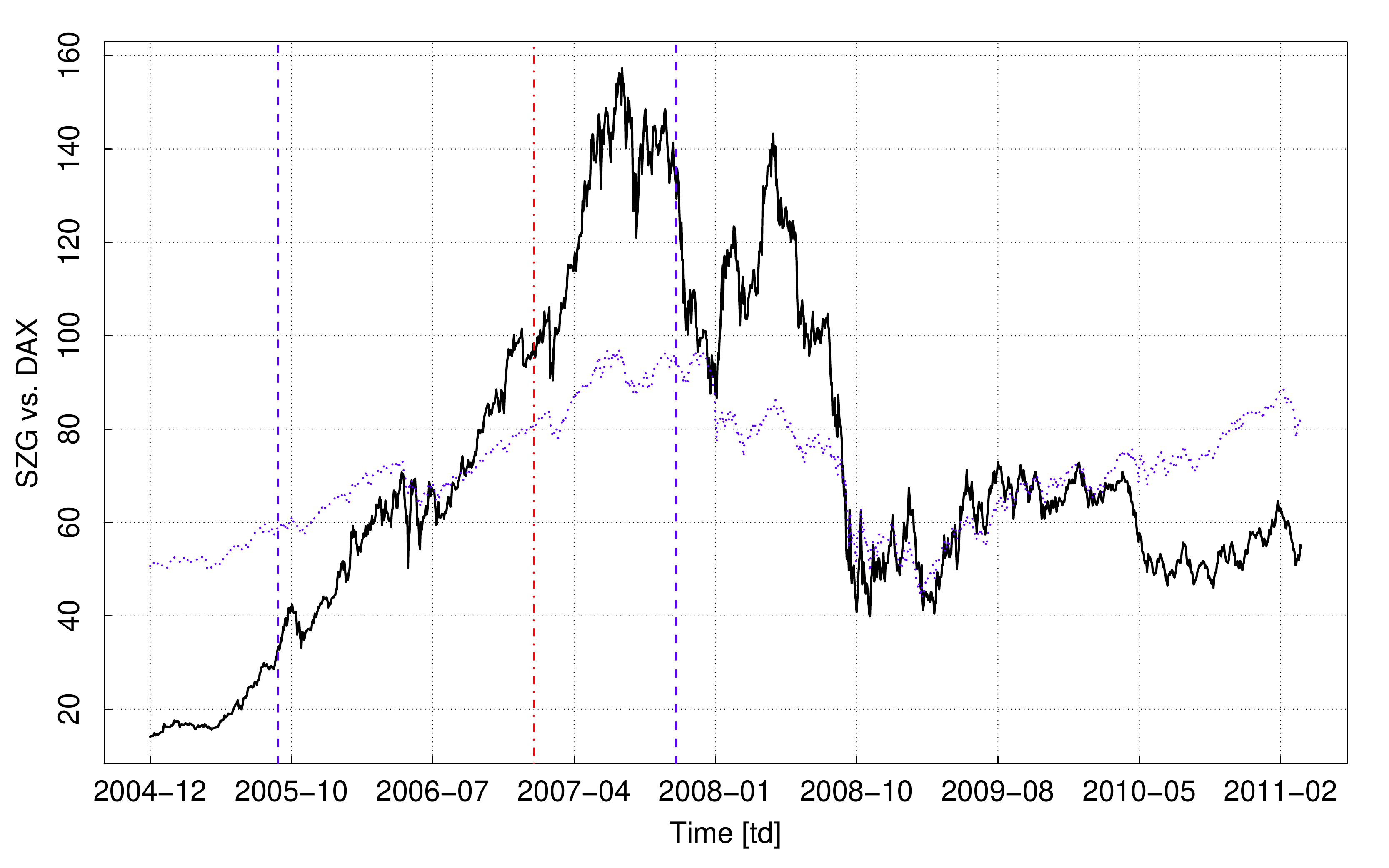}
\caption{Quotations of the SZG company (black solid curve) together with the rescaled DAX (blue dashed curve) ranging from 2004-12-27 (Monday) to 2011-03-25 (Friday). The scaling factor of the DAX is given  by the ratio of the mean value of the SZG series to the mean value of the DAX series. This resulted in a better separation of the plots. The roles of vertical dashed (blue) and dashed-dotted (red) lines are described in the main text.}
\label{figure:DAX_SZG}
\end{center}
\end{figure}

We take into account, as a representative example of complex network dynamics, the evolution -- in daily and weekly horizons -- of the most liquid number of survived companies, $n=459$, quoted on the Frankfurt Stock Exchange\footnote{For comparison, the DAX contains only 30 largest companies.} (FSE) during the particularly significant and intriguing period of large FSE variability. Full time series which we use begins at 2004-03-22 (Monday) and finishes at 2011-12-30 (Friday). However, since we use the optimal scanning window of 400 trading days width, the centre of the scanning window scans a shorter time series -- from 2004-12-27 (Monday) to 2011-03-25 (Friday), still containing the recent worldwide financial market crisis and crash (for details, please refer to the quotation plots of the SALZGITTER (SZG) AG-Stahl und Technologie company and the DAX in Fig. \ref{figure:DAX_SZG}, as well as the underlying plots in our earlier paper \cite{WSGKS}). These $459$ survived companies define the FSE basis -- hence, we can consider the total numbers of MST vertices and edges as conserved (non-fluctuating) quantities. 

Seemingly, the methodology used in this work and based on conserved quantities, appeared to be quite different from that used in our earlier works 
\cite{WSGKS,SGKS}, where number of the most liquid companies fluctuate by a few percent from time to time. Nevertheless, the results obtained with both methodologies are essentially indistinguishable within the assumed resolution, from those provided in the figures contained in this work. More precisely, we only observed that the absolute MOL's minimum is now located at 2005-01-25 (Thursday) instead of two trading days later, i.e. at 2007-01-29 (Monday). Presumably, this small shift is due to sensitivity of the MOL which is much greater than the mean tree length \cite{RNM,BLM,BR,TSC,WSGKS,SGKS}. It is worth noting that, the logarithm of the mean occupation layer resembles nonequilibrium entropy of a complex network (compare plots presented in Figs. 
\ref{figure:entropy_deg_eff} and \ref{figure:MOLs}), which helps to identify the temporal key vertices in the complex network \cite{JulReh}.

\subsection{Initial empirical evidences}\label{section:Dpaed}

\begin{figure}
\begin{center}
\bigskip
\includegraphics[width=160mm,angle=0,clip]{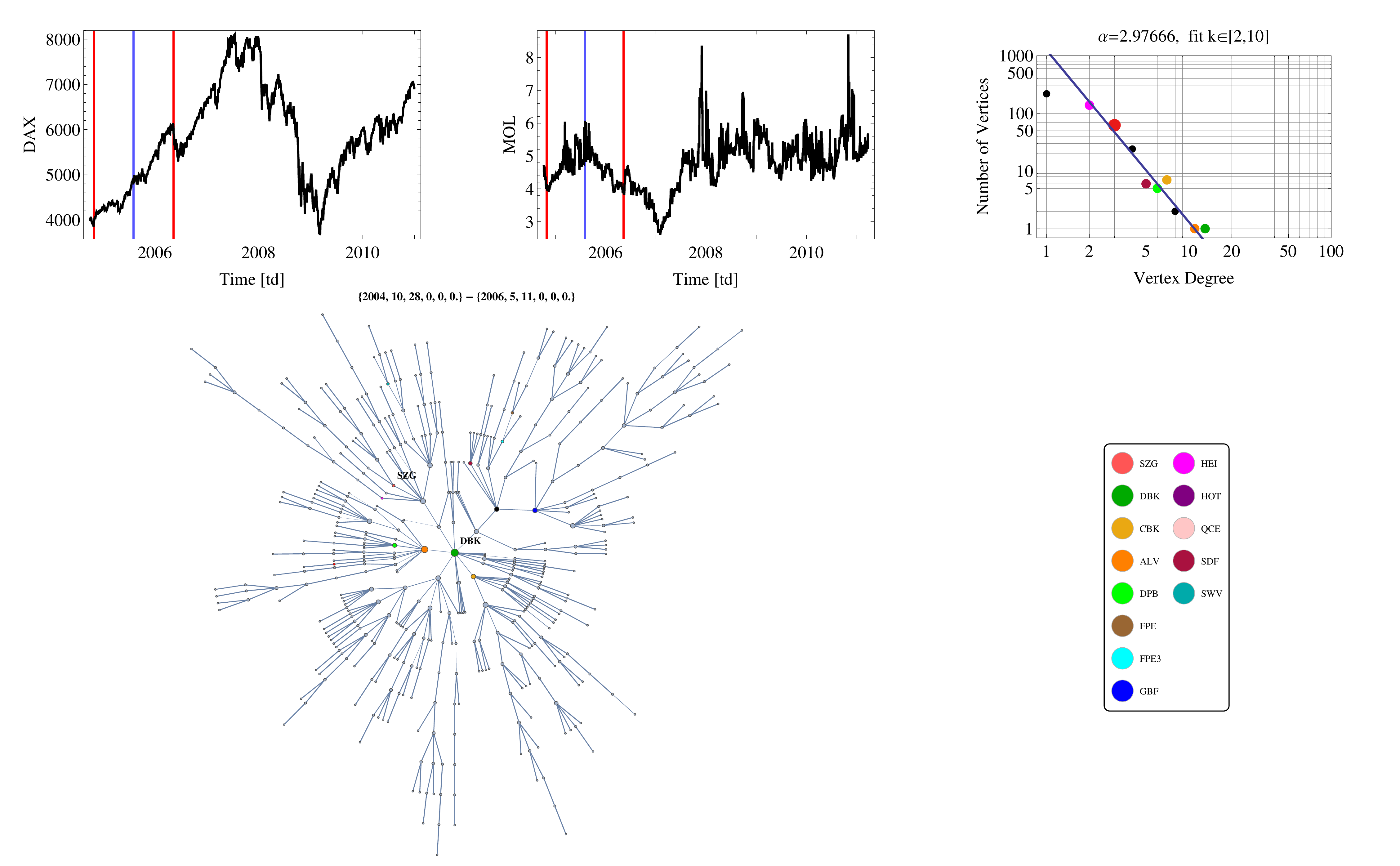}
\caption{Snap-shot picture of the empirical MST (graph placed in the lower row on the left-hand side of the figure) consisting of companies quoted on the FSE within the sub-period from 2004-10-28 (Thursday) to 2006-05-11 (Thursday) -- these boundary dates are denoted by the red vertical straight lines in two plots placed in the upper row. The center of this sub-period -- at 2005-08-03 (Wednesday) -- is denoted by the blue vertical straight line (defining the frame of our movie no. 16). The plot in log-log scale, placed in the upper row on the right-hand side of the figure, shows the fit of (unnormalized) empirical distribution of vertex degrees (small circles) by a power law (solid sloped straight line). Notably, the boundary companies (having degree equal 1) are coming off the power law (in all snap-shot pictures) as a result of a finite size effect. The colored points in this plot represent not only the corresponding companies but also others of the same vertex degree. The name abbreviations of thirteen most significant companies (marked by the color circles -- the same in the plot and graph) are listed in the legend (placed in the right lower corner of the figure). 
Since we use the scanning window of 400 trading days width, the full time series which we use, is longer than that given above as it begins at 2004-03-22 (Monday) and finishes at 2011-12-30 (Friday). Therefore, the centre of the scanning window scans a shorter time series, i.e. from 2004-12-27 (Monday) to 2011-03-25.
}
\label{figure:ZE_16}
\end{center}
\end{figure}

\begin{figure}
\begin{center}
\bigskip
\includegraphics[width=160mm,angle=0,clip]{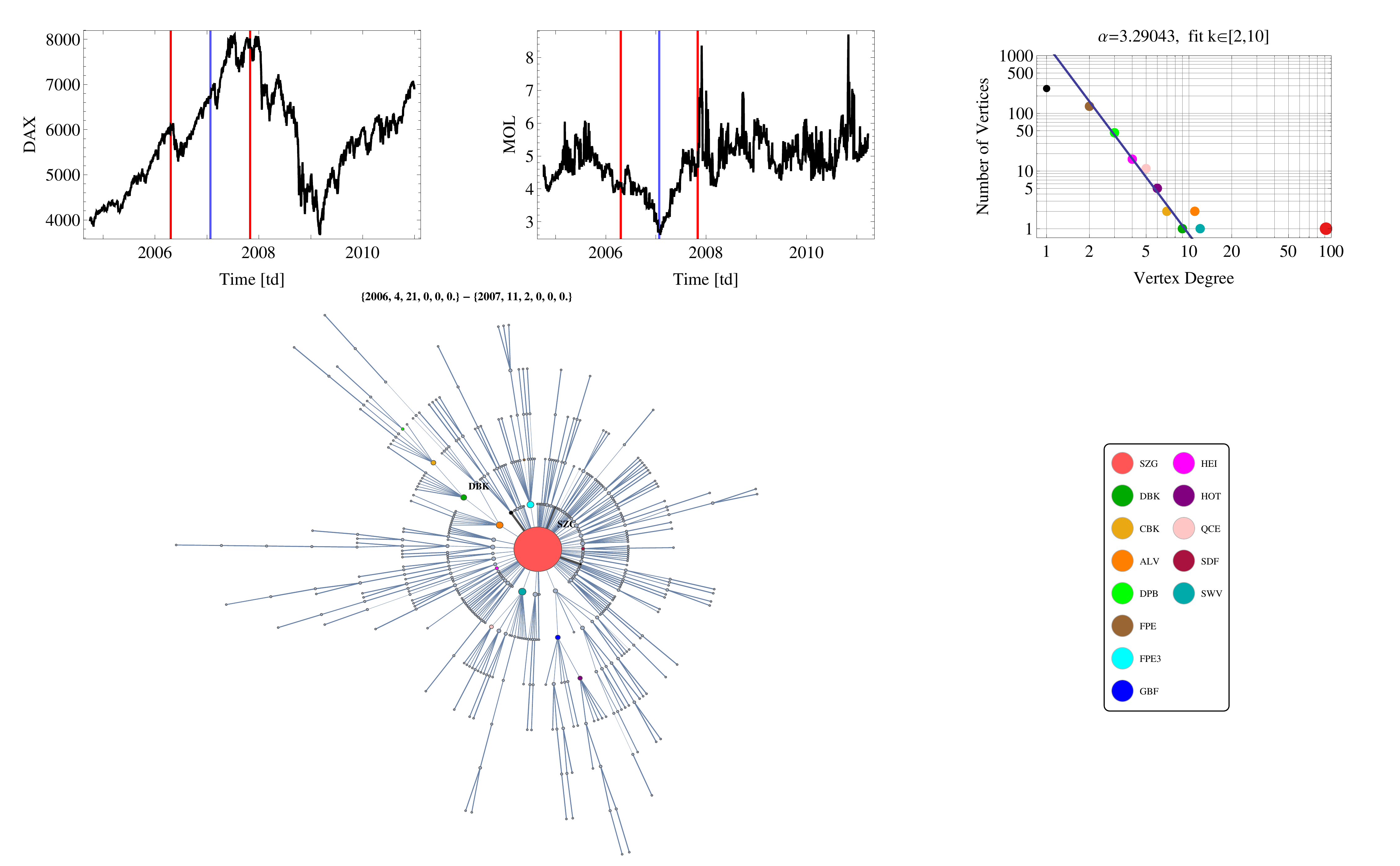}
\caption{Snap-shot picture of the empirical MST (graph placed in the lower row on the left-hand side of the figure) consisting of the most liquid companies quoted on the FSE within the sub-period from 2006-04-21 (Friday) to 2007-11-02 (Friday) -- these boundary dates are denoted by the red vertical straight lines in both plots placed in the upper row. The center of this sub-period -- at 2007-01-29 (Monday; the frame of our movie no. 402) -- is denoted by the blue vertical straight line. It is worth noting, how much the SZG company is now coming off the power law -- its degree, equals 91, which is a maximal value reached by the SZG vertex during its evolution. This number was calculated using the methodology which considers a fluctuating number of the most liquid companies quoted on the FSE. For another methodology, considering a fixed number of companies, this number equals 88 and the corresponding centre of the sub-period is located at 2007-01-25 (Thursday), which is indistinguishable (within assumed resolution of the Figure) from date 2007-01-29 (Monday) given above. This difference has no noticeable influence on any of our results therefore, we use both. Besides, SWV, ALV, and FPE3 companies (occupying the second and \emph{ex aequo} the third positions in the rank, respectively) are slightly coming off this power law. These companies were already `attracted' by SZG vertex to its first and second coordination layers (zones). Furthermore, DBK (the initial leader, when the MST network was in the `Equilibrium 
scale-invariant network') occupies now the fourth position in the rank and is located in the second coordination zone. In fact, all these companies are located now much more closely to the SZG vertex than earlier, i.e. during the nucleation process \cite{KGSH,KSzW1,KSzW2} 
(see Figs. \ref{figure:ZE_134} and \ref{figure:ZEP_341} for details).}
\label{figure:ZEP_402}
\end{center}
\end{figure}

We apply the MST technique to investigate transient behavior of a complex network during its evolution from 
a scale-free topology representing the initial equilibrium stock market hierarchical structure, long before the recent worldwide financial crash 
\cite{DiSo} -- its typical structure is presented in Fig. \ref{figure:ZE_16} -- to the one dominated by superstar-like tree (superhub or dragon king presented in Fig. \ref{figure:ZEP_402} by the red large central circle) resembling a dissipative structure \cite{IlPrig} which, after adaption to external conditions, decays if these conditions disappear.
Apparently, this superhub decorated by scale-invariant hierarchy of trees placed in its first, second and further coordination zones (or occupation layers), represents the market structure during the period containing the crash. 

\begin{figure}
\begin{center}
\bigskip
\includegraphics[width=160mm,angle=0,clip]{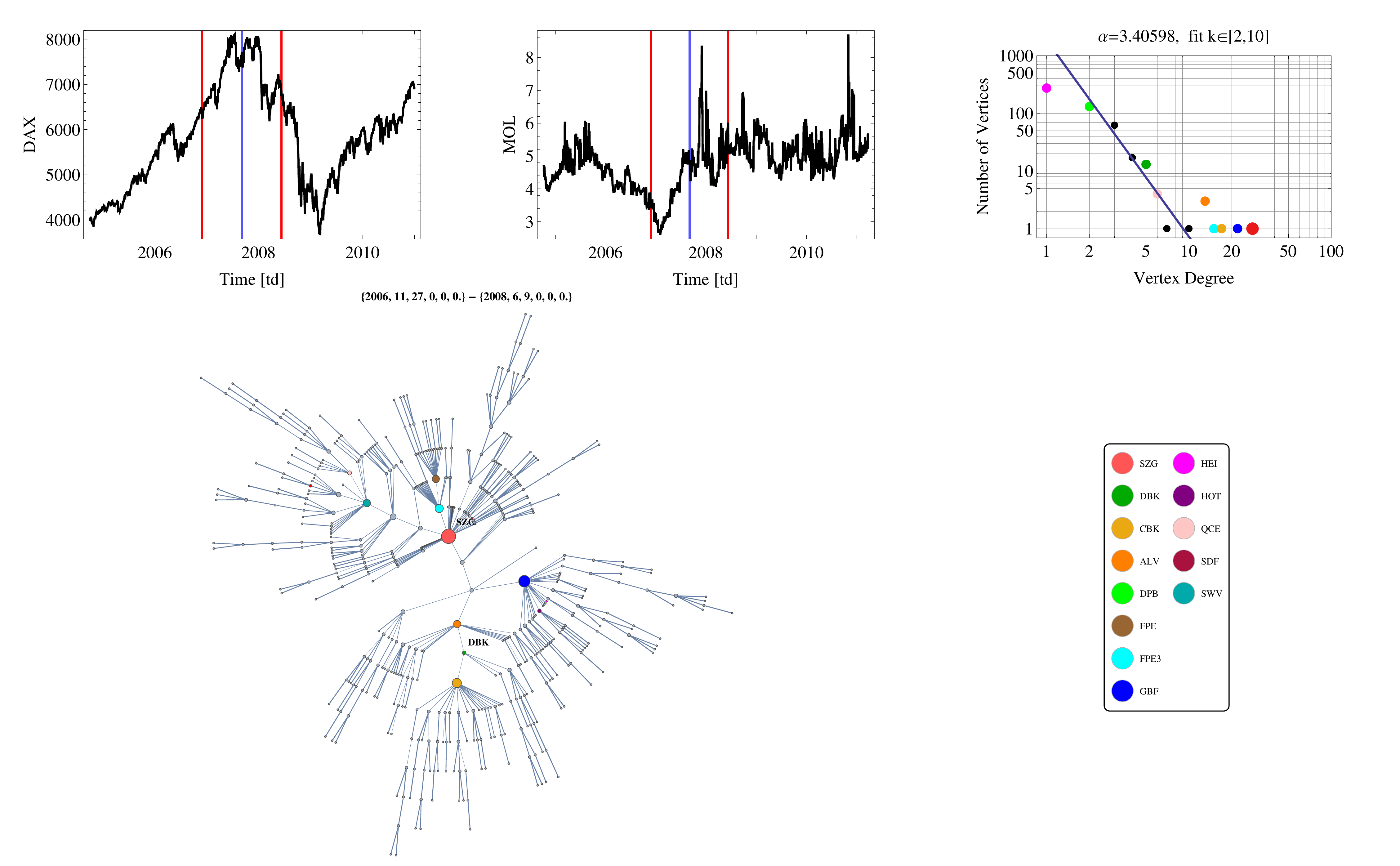}
\caption{Snap-shot picture of the empirical MST (graph placed in the lower row on the left-hand side of the figure) consisting of companies quoted on the FSE within the sub-period from 2006-11-27 (Monday) to 2008-06-09 (Monday) -- these boundary dates are denoted by the red vertical straight lines in both plots placed in the upper row. The center of this sub-period -- at 2007-09-03 (Monday; the frame of the movie no. 558) -- is denoted by the blue vertical straight line. Now, the strong rearrangement of the MST network is well seen, e.g., in comparison with the corresponding one presented in Fig. 
\ref{figure:ZEP_402}, as seven vertices are coming off the power law, although, the SZG company still occupies the leader position. However, this is already with a little predominance -- now, its vertex degree equals only 28, while vice-leader, GBF company, has degree equals 22. Apparently, the current MST network is strongly decentralized, consisting of several (at the moment of seven) well distinguished clusters. Indeed, such a structure is typical for the MST network after the entropies, MHSD, and MOL reach their respective minima (see Figs. \ref{figure:entropy_deg_eff} -- \ref{figure:MOLs}), where during the time lag the leader position is exchangeably occupied by different companies.}
\label{figure:ZEP_558}
\end{center}
\end{figure}

Furthermore, the complex networks presented in Figs. \ref{figure:ZE_16} -- \ref{figure:ZEP_558} have essentially different but quite typical modular structures. The central element of the structure presented in Fig. \ref{figure:ZE_16} is the two-node core (consisting of DBK and ALV companies of degrees greater than ten). The structure presented in Fig. \ref{figure:ZEP_402} is the superstar-like one centered at the SZG vertex (having degree greater than ninety, i.e. much greater than degrees of all other nodes), which later takes the modular structure shown in Fig. \ref{figure:ZEP_558}. This latter structure consists of two well-separated parts: the first, single-core one centered at the SZG node (having degree equals twenty eight) and the second, three-core structure, where separated cores are centered at GBF, CBK, and ALV nodes (of degrees greater than fifteen). However, this is with only a little SZG predominance -- the vice-leader, GBF company, has already degrees equals twenty two. More precisely, the current MST network is strongly decentralized, consisting of several (at the moment of seven) well distinguished clusters. Indeed, such a structure is typical for the MST network after the entropies, MHSD and MOL reach their respective minima (see for details Figs. \ref{figure:entropy_deg_eff} -- \ref{figure:MOLs}, respectively) still placed at the same day, while during the time lag the leader position is exchangeably occupied by different companies.  

Notably, our movie and hence all snap-shot pictures were calculated using the methodology which considers a fluctuating number of the most liquid companies quoted on the FSE. For another methodology, considering a fixed number of companies, the maximal vertex degree equals 88 instead of 91 and the corresponding centre of the sub-period is located at 2007-01-25 (Thursday), which is indistinguishable (within assumed resolution of figures) from date 2007-01-29 (Monday) given above. This difference has no noticeable influence on any of our results therefore, we use both.

The structural and topological phase transitions, similar to those considered in the present work, we also found on the Warsaw Stock Exchange (WSE) -- a complex network of 274 companies, quoted on the WSE throughout the period in question. Here, we omitted the results obtained for the WSE because they resemble those found for the FSE and they have been presented in Ref. \cite{SGKS}. Both our results, that is for WSE and FSE, are complementary to those found by Onnela, Chakraborti, Kaski, and Kert\'esz for 116 stocks of S\&P 500 index in the vicinity of Black Monday (October 19, 1987) \cite{OCKK1} and also in the vicinity of January 1, 1998 \cite{OCKK2}. However, all of them suggest: (i) very significant role of the MOL in study of the MST evolving structure and (ii) the appearance of a distinct absolute minimum of the MOL, which can be considered to be a significant precursor of a crash. Indeed, the thrilling consequence of our work would be practically to verify this conjecture and its far-reaching consequences. Here we theoretically consider \emph{a posteriori}, the illuminating and surprising, yet universal emergent properties of the continuous phase transition from the equilibrium scale-free MST network to the non-equilibrium nucleation preceding the condensation of edges forming a $\lambda $-transition. Notably, the critical instant $t_{\lambda }$, falls on the date 2007-01-25 (Thursday) when number of vertices is fixed otherwise, it corresponds to the date 2007-01-29 (Monday).

The main goal of our work is to show, that the superhub forms a temporal structural condensate on a real-life financial market. Subsequently, we aim to present the transition of the $\lambda $-type from this condensate to scale-invariant topology decorated by the hierarchy of local star-like hubs, representing the market structure and topology directly after the worldwide financial crash. To observe and analyze the above mentioned temporal structure of evolving complex network we use sufficiently sensitive characteristics, for instance, the order parameters presented in Figs. 
\ref{figure:entropy_deg_eff} -- \ref{figure:MOLs} and others considered below.

\subsection{Dominant role of the SZG company}\label{section:DrSZG}

It is illustrative to document, as a typical reference example, that the SALZGITTER (SZG) AG-Stahl und Technologie company becomes a dominant node of the Frankfurt Stock Exchange MST network during the worldwide financial and economical crisis, still persisting to date. 

\begin{figure}
\begin{center}
\bigskip
\includegraphics[width=140mm,angle=0,clip]{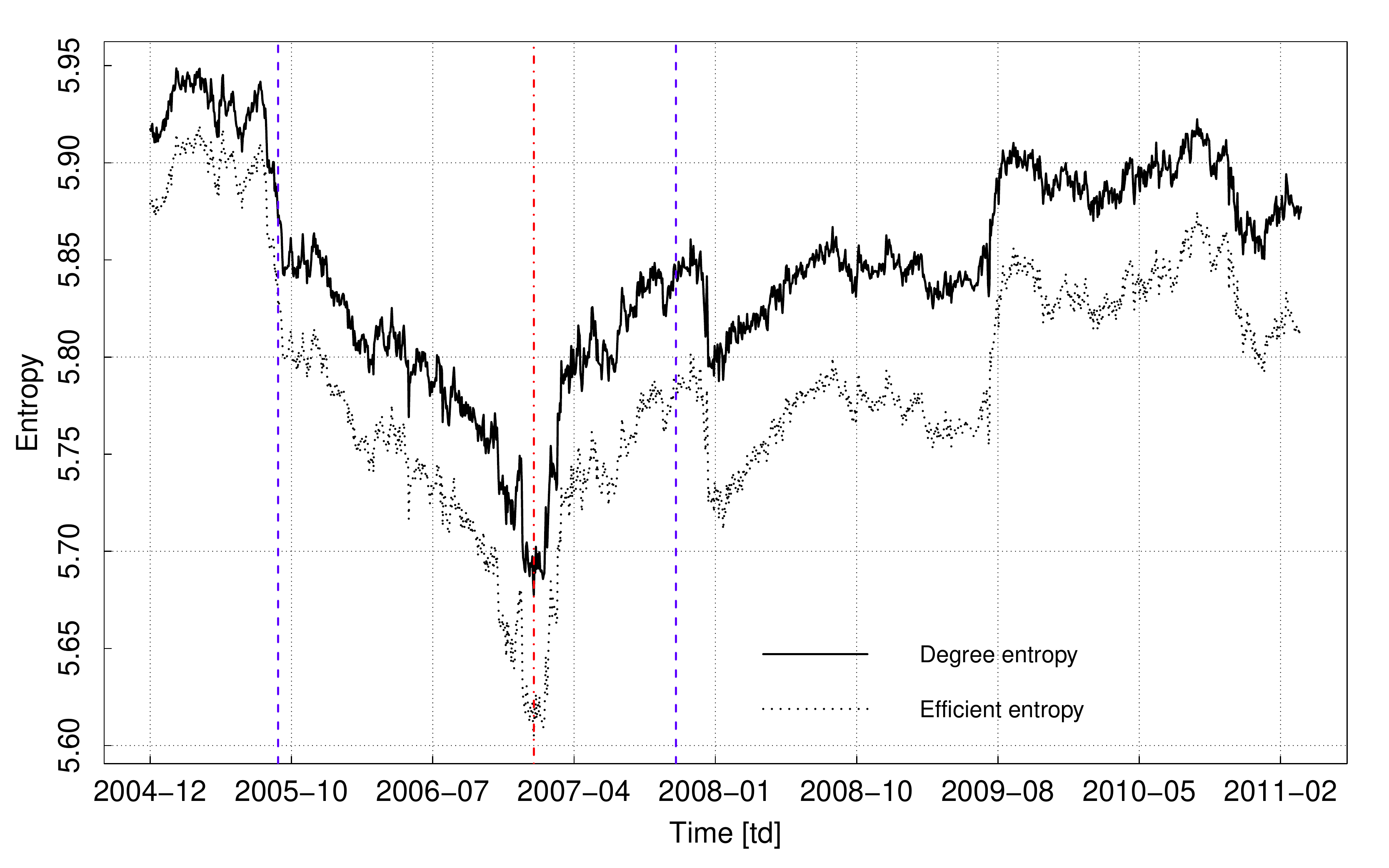}
\caption{Two different dynamical non-equilibrium entropies for the MST network, derived for the FSE (and counted in trading days (td)). That is, the plot of the time-dependent degree (solid curve), and efficiency (dashed curve) entropies (for the entropy definitions see Ref. \cite{WSGKS}). The red dashed-dotted vertical line is located at the common absolute minimum of both entropies, i.e. at 2007-01-25 (Thursday). The role of the blue dashed vertical lines was clarified in the main text and illustrated in Fig. \ref{figure:diff_mol}.}
\label{figure:entropy_deg_eff}
\end{center}
\end{figure}

\begin{figure}
\begin{center}
\bigskip
\includegraphics[width=170mm,angle=0,clip]{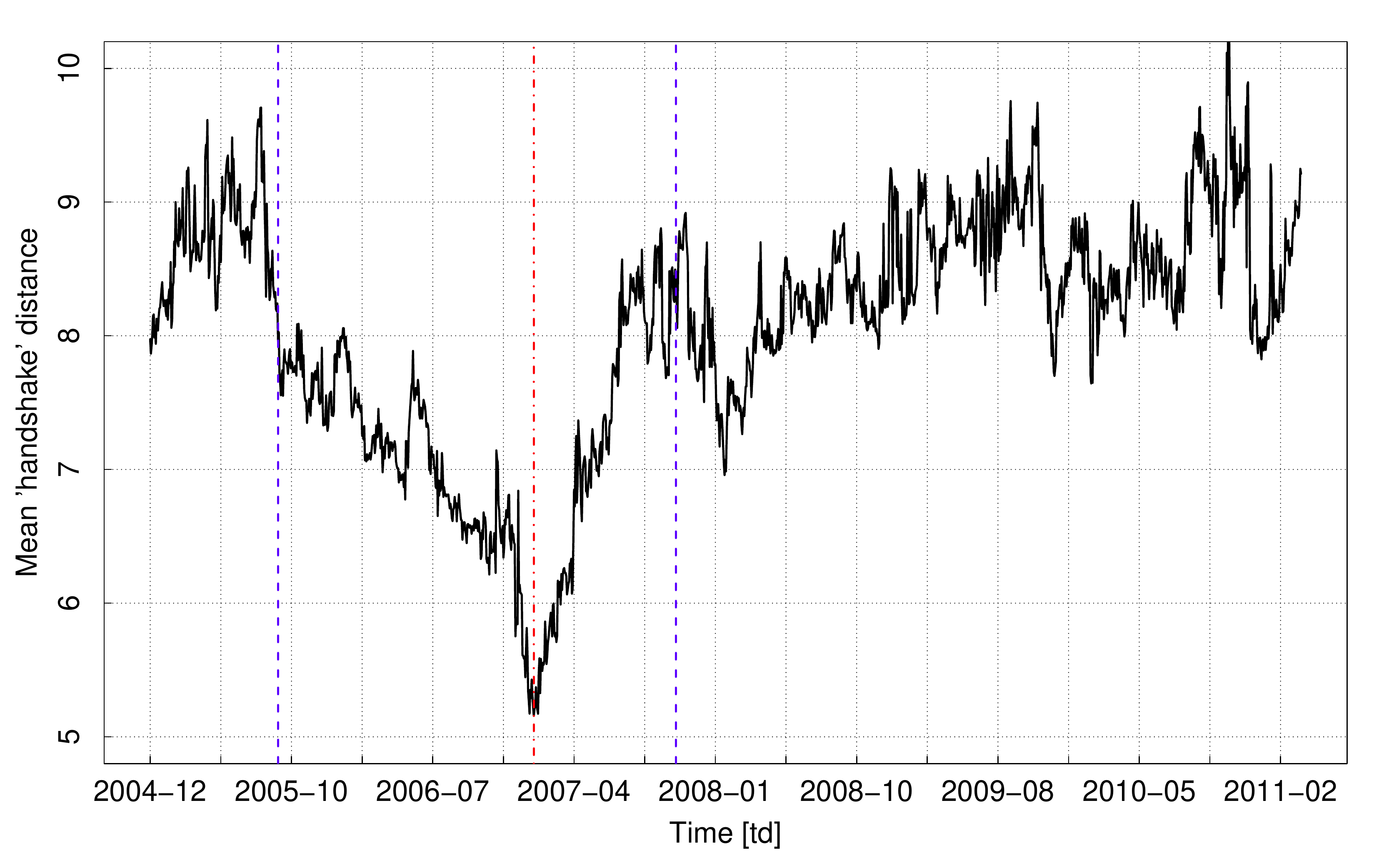}
\caption{Mean `Handshake' Distance (MHSD) calculated for the MST network vs. time. The internode distance, MHSD, between any pair of vertices is measured by the corresponding number of edges forming a unique path connecting these vertices. The absolute minimum (at 2007-01-25 (Thursday)) was denoted by the red dashed-dotted vertical line. Remarkably, this absolute minimum and others, shown in Figs. \ref{figure:entropy_deg_eff} and \ref{figure:MOLs}, coincide. In the surroundings of this minimum, the MST network can be considered as compact and `small world'-like (because then MHSD$ \propto \ln n$). The role of the blue vertical dashed lines was clarified in the main text and illustrated in Fig. \ref{figure:diff_mol}.}
\label{figure:smallw}
\end{center}
\end{figure}

\begin{figure}
\begin{center}
\bigskip
\includegraphics[width=140mm,angle=0,clip]{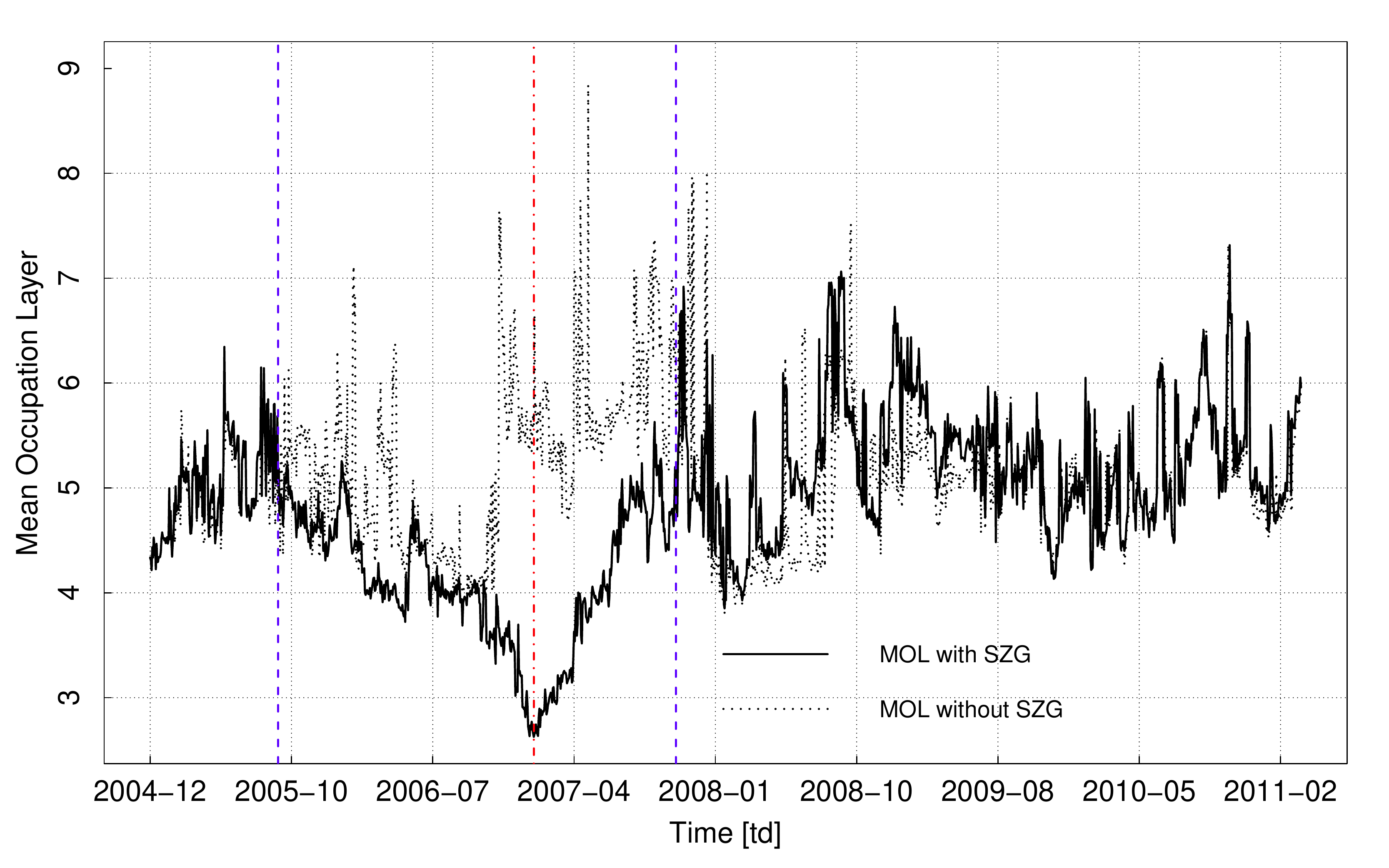}
\caption{Mean Occupation Layer (MOL) calculated for two different MST networks. Namely, for the first one containing the SZG vertex (solid curve) and for the second one which does not contain SZG (dashed curve). The essential difference between both various MOLs is a relatively deep absolute minimum well visible for the plot without the SZG (at 2007-01-25 (Thursday) denoted by the red dashed-dotted vertical line). Remarkably, this absolute minimum and others, shown in Fig. \ref{figure:entropy_deg_eff}, are located at the same trading day. The role of the blue vertical dashed lines (the same as plotted in Fig. \ref{figure:diff_mol}) was explained in the main text and illustrated in Fig. \ref{figure:diff_mol}.}
\label{figure:MOLs}
\end{center}
\end{figure}

\begin{figure}
\begin{center}
\bigskip
\includegraphics[width=140mm,angle=0,clip]{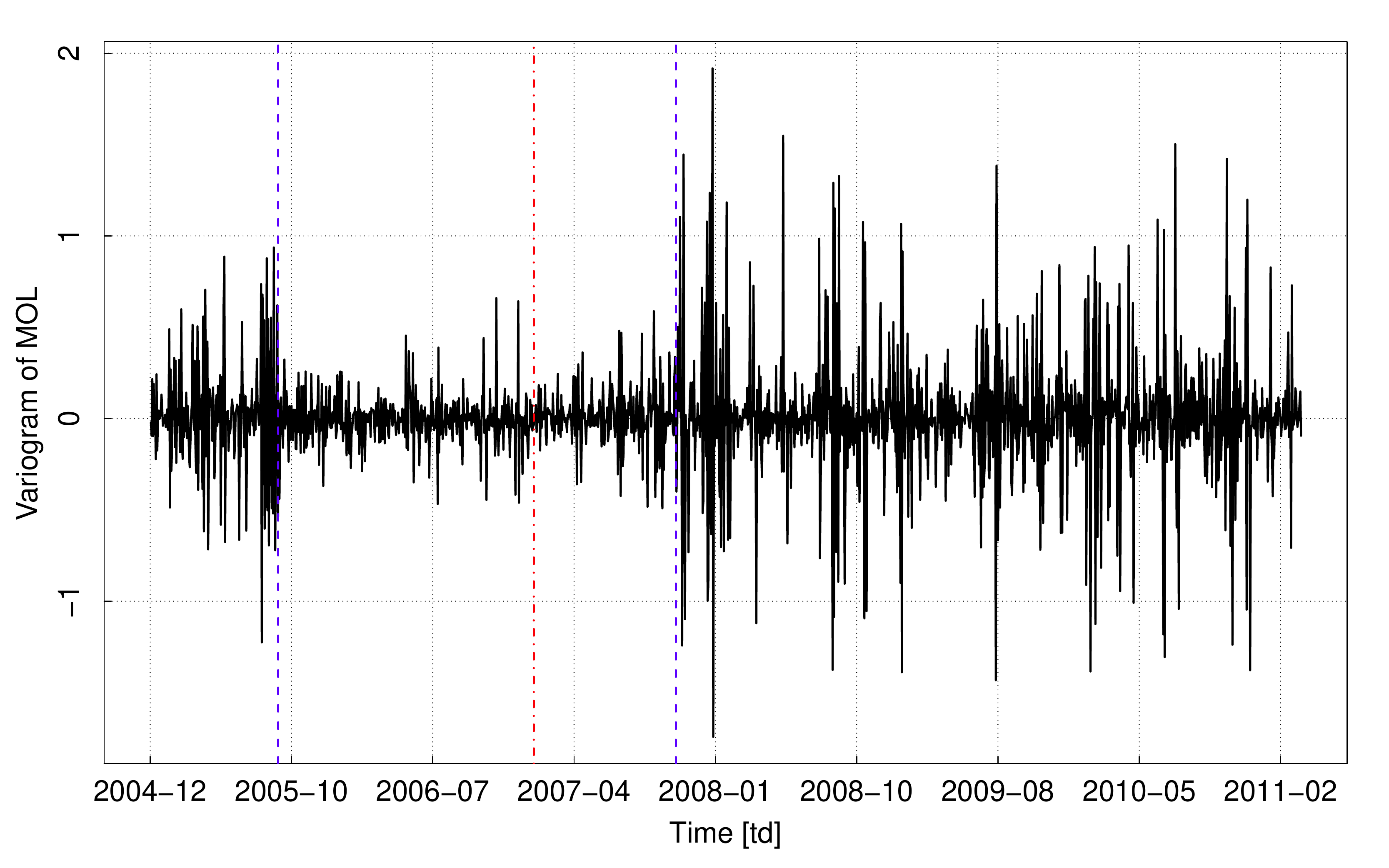}
\caption{Variogram of the Mean Occupation Layer (MOL) calculated for the MST network containing the SZG vertex (see Fig. \ref{figure:MOLs}). Significant diminishing of its amplitude (within the sub-period restricted by the blue dashed vertical lines) is well seen, in particular, in the vicinity of the MOL's absolute minimum (i.e. at 2007-01-25 (Thursday) marked by the dashed-dotted vertical line). In other words, blue dashed vertical lines define the time range from 2005-09-16 (Friday) to 2007-12-14 (Friday), where amplitude of the variogram is distinctly lower than for the outer regions. This behavior resembles that well known in the presence of a fixed point.}
\label{figure:diff_mol}
\end{center}
\end{figure}

\begin{figure}
\begin{center}
\bigskip
\includegraphics[width=140mm,angle=0,clip]{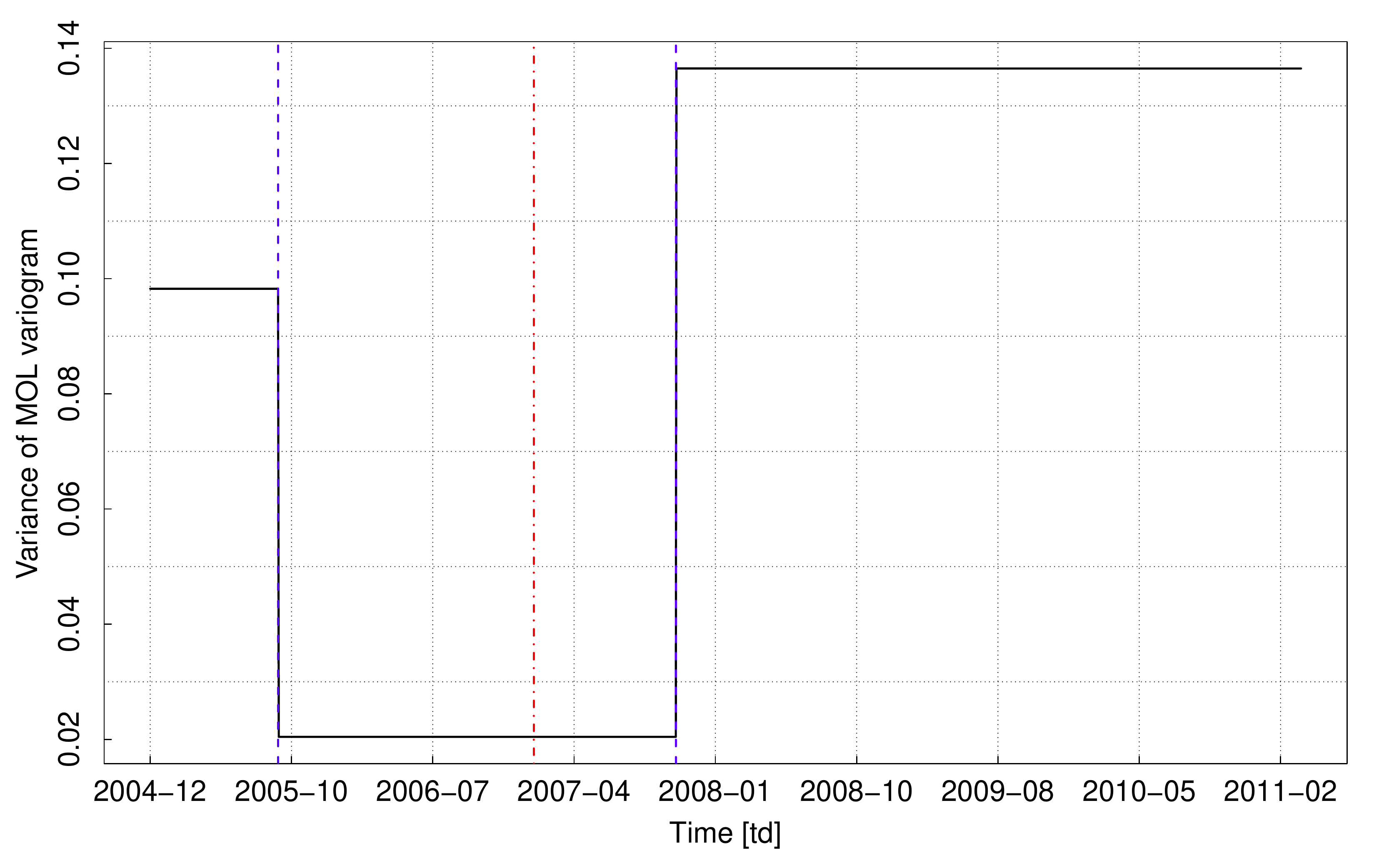}
\caption{The plot of (partial) variances of the MOL variogram (see Fig. \ref{figure:diff_mol}) calculated for the MST network containing the SZG vertex (see Fig. \ref{figure:MOLs}). The expected significant diminishing of the variance is well seen in the surrounding (restricted by the blue dashed vertical lines) of the MOL's absolute minimum (at 2007-01-25 (Thursday) marked by the red dashed-dotted vertical line). Apparently, there is a strong increase of the variance outside this surrounding, in particular for the longer time. Inside this surrounding, the MST network structure is sufficiently robust against fluctuations in comparison with the remaining two MST network structures. Also this behavior resembles that in the presence of a fixed point.}
\label{figure:diff_mol_var}
\end{center}
\end{figure}

We study the evolution of the MST network before and after the absolute minimum of the time-dependent entropies 
\cite{HSPV} plotted in Fig. \ref{figure:entropy_deg_eff}. Although both plots look very similar there, the definitions of the corresponding entropies are quite different (cf. Ref. \cite{WSGKS}) -- it is unnecessary to exploit this point in the present work.
Apparently, both entropies (Fig. \ref{figure:entropy_deg_eff}), Mean `Handshake' Distance (MHSD; see plot in Fig. \ref{figure:smallw}), and MOL (plotted in Fig. 
\ref{figure:MOLs} with the solid curve) have quite similar shapes and, the most significant observation, is their absolute minima coincide at Thursday 2007-01-25 (cf. also our earlier considerations given in Refs. \cite{WSGKS} and \cite{SGKS}).  
We can speculate, that at this minimum the least disordered state (cf. Fig. \ref{figure:ZEP_402}) of the MST network is located, just between the preceding (cf. Fig. \ref{figure:ZE_16}) and following (cf. Fig. \ref{figure:ZEP_558}) more disordered states. These more disordered states are located outside the region limited by the blue vertical dashed lines, where the amplitude of the MOL variogram shown in Fig. \ref{figure:diff_mol}, is distinctly higher than inside this region. 
This observation is well confirmed by the behavior of the corresponding (partial) variances of MOL variogram presented in Fig. \ref{figure:diff_mol_var}.  
Such a behavior is typical for random variable remaining a longer time in the surroundings of a stable fixed point. 

\begin{figure}
\begin{center}
\bigskip
\includegraphics[width=140mm,angle=0,clip]{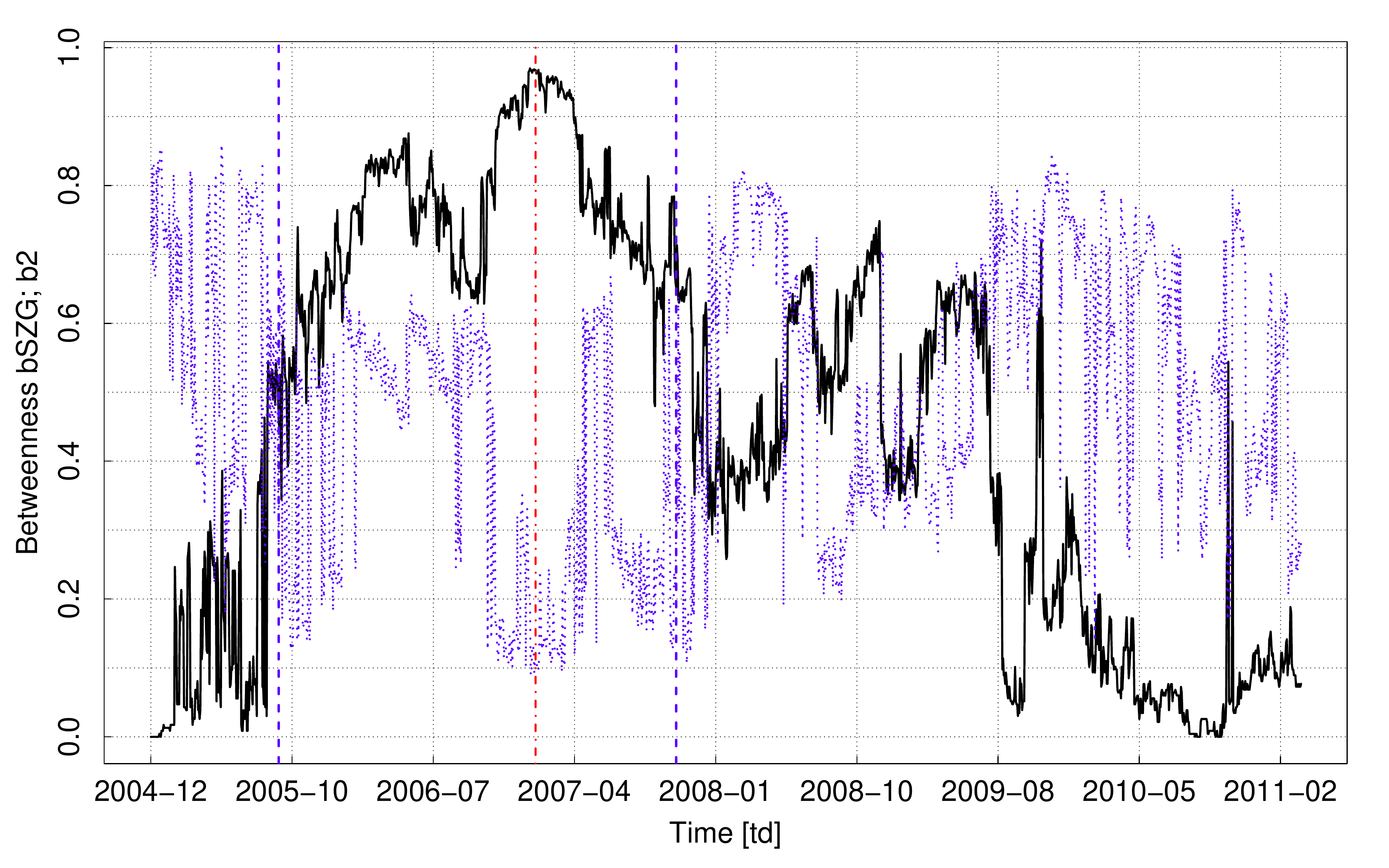}
\caption{Temporal betweenness \cite{PB,MPA,PMA} of the SZG company (denoted by the black erratic solid curve) and betweenness of the vice-leader (the blue erratic dashed curve). The vice-leader position is not occupied all the time by the same company. The occupation of this position varies from time to time. Remarkably, the leader position is occupied by the SZG company only when its betweenness is greater than the one of the temporal vice-leader. It takes place for the time interval limited by the blue dashed vertical lines -- the same, as shown in Fig. \ref{figure:diff_mol}. Apparently, the betweenness of the SZG reaches the absolute maximum at 2007-01-25 (denoted by the red dashed-dotted vertical line) that is, it reaches this maximum at the same day when, for instance, MOL takes the absolute minimum (cf. Fig. \ref{figure:MOLs}). That day the betweenness of the vice-leader (it is then the SWV holding company) also assumes an absolute minimum.}
\label{figure:betweeness}
\end{center}
\end{figure}

The significance of a central role of the SZG company within the sub-period from 2005-09-16 (Friday) to 2007-12-14 (Friday) (limited by blue dashed vertical lines plotted, for instance, in Figs. \ref{figure:DAX_SZG} and \ref{figure:entropy_deg_eff} -- \ref{figure:betweeness}) is well captured by the conformity of the two temporal (simplified) betweennesses, $b_{SZG}$ and $b_2$ (defined by Eq. (137) in Ref. \cite{KwDr}), shown in Fig. 
\ref{figure:betweeness}. 
Apparently, within the central peak (located around January 25, 2007 {as its center}), the number of paths passing through the SZG vertex is about seven times larger than those passing through the vice-leader vertex, where the vice-leader vertex is defined as occupying the second position in the rank of vertex degrees. Here, it is played mainly by the SWV holding company\footnote{SolarWorld (SWV) AG holding company is engaged in the production of the crystalline solar power technologies.} being the leader of the different `Sector of Renewable Energy Equipment'. Indeed, the role of the SZG company is substantially greater than that of the SWV one. 

The evidence is given in Section \ref{section:beftlamb} why our study was formally inspired by properties of $ ^4$He,  which  below ${\lambda }$-line is in a superfluid II $ ^4$He phase, while in the normal fluid I $ ^4$He phase only above \cite{RNS}. Comments concerning $\lambda $-transition of $ ^4$He to the superfluid phase and its hypothetical relation to the Bose-Einstein condensation, can be found in Ref. \cite{KHuang,PGDSz}. This correspondence has only a formal character because in our study the role analogous to the inverted temperature is played by time, that is, we deal with a dynamical phase transition and not with the thermodynamic one.

\subsection{Empirical evidences for nucleation, condensation and $\lambda $-transition}\label{section:beftlamb}

This paragraph contains our detailed considerations concerning the most intriguing part of the MST network evolution, which occurs for time 
$t\leq t_{\lambda }$.
 
\begin{figure}
\begin{center}
\bigskip
\includegraphics[width=140mm,angle=0,clip]{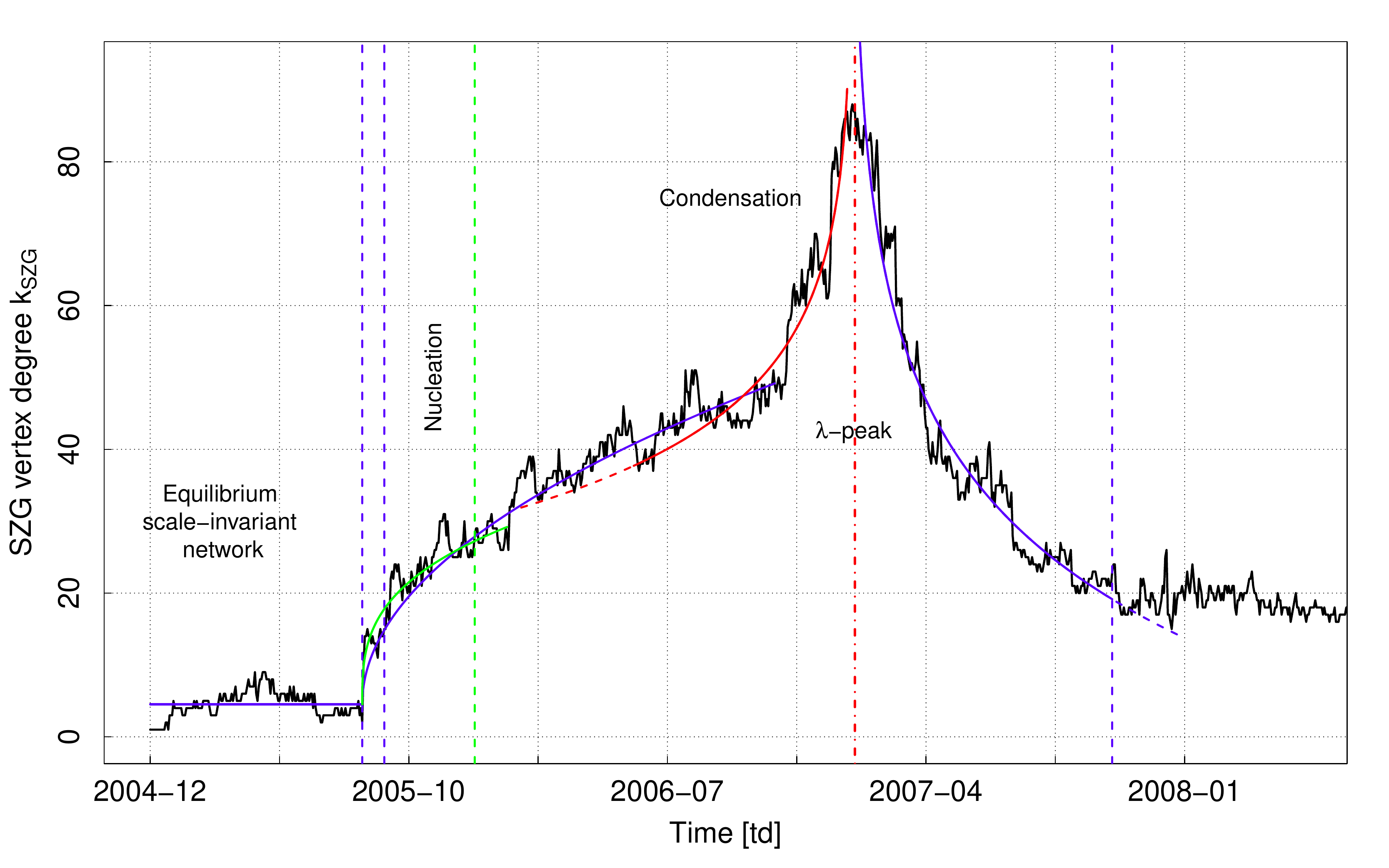} 
\caption{The plot of the temporal SZG vertex degree, $k_{SZG}$, which forms, striking the eye, $\lambda $-peak centered at 
$t_{\lambda }=t_{min}^{MOL}=544$[td]$\equiv 2007-01-25$ (Thursday) (which site is denoted by the red dashed-dotted vertical line). Herein, time plays the role of the control parameter. The empirical data are shown by the black erratic solid curve obtained for the daily horizon. The blue solid curve, consists of three parts (but the latter part is separated by the red one). The height of the first horizontal part equals $A_0=4.5273$ -- a mean value of $k_{SZG}$ before the first transition time or the first critical threshold $t_{crit}= 164$[td]$\equiv ${2005-08-11 (Thursday)} (the location of this threshold is denoted by the first blue vertical dashed line). The second long-term part (of the order of one year), beginning at $t_{crit}$, is described by a power law function $A (t- t_{crit})^{1/z}+A_0$, where the global dynamic exponent $z=2$ and amplitude $A=2.50$. However, the early stage of the semi-critical dynamics (of the order of one month, ranging between two subsequent blue dashed vertical lines) is driven by the canonical Lifshitz-Slyozov dynamic exponent $z=3$ and $A=5.20$ (the green curve). The location of the green vertical dashed line is defined further in Fig. \ref{figure:kondensat_day}. The third part of the blue solid curve is defined for $0 < t - t_{\lambda } < \tau _R$ by the logarithmic relaxation function 
$-A_R\ln \left(\left(t-t_{\lambda }\right)/\tau _R\right)$, where $A_R=22$ and $\tau _R=480$[td]. This function is a solution (\ref{rown:sollambdaR}) of the macroscopic Eq. (\ref{rown:lambdaR}). The red impetuously increasing solid curve represents a logarithmic function 
$-A_L\ln \left(\left(t_{\lambda }-t\right)/\tau _L\right)$, for $0<t_{\lambda }-t<\tau _L$, where amplitude $A_L=14$ and $\tau _L=2500$[td]. The $k_{SZG}$ short-range cross-over (of the order of one quarter) from nucleation to condensation is defined somewhere in the surroundings of 2006-07 by the overlap of blue and red solid curves, where no sharp transition is observed.}
\label{figure:oba_zbocza}
\end{center}
\end{figure}

\begin{figure}
\begin{center}
\bigskip
\includegraphics[width=140mm,angle=0,clip]{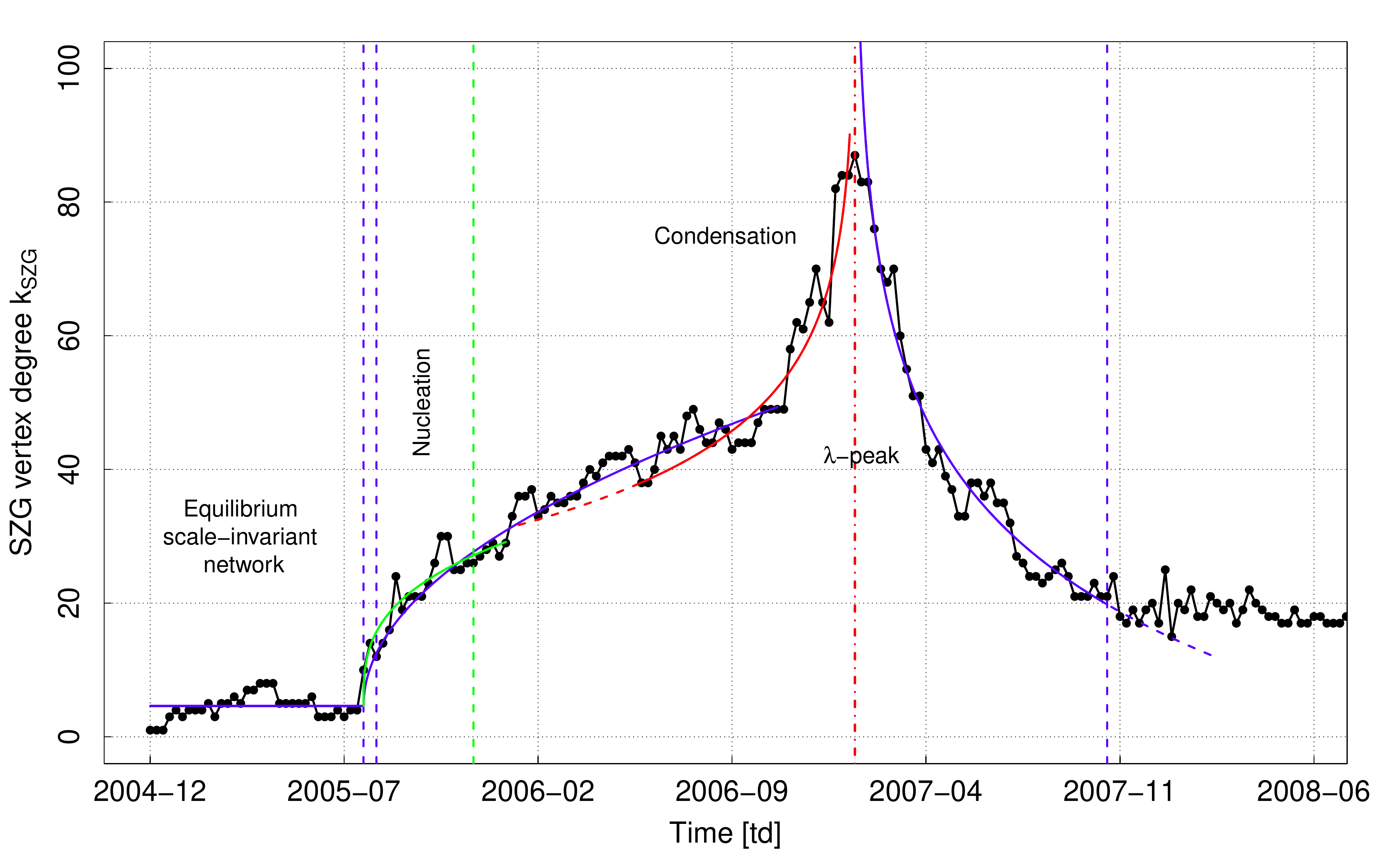} 
\caption{Striking the eye plot of the SZG vertex degree, $k_{SZG}$, vs. time forming the $\lambda $-peak centered at 
$t_{\lambda }=t_{min}^{MOL}=544$[td]$\equiv 2007-01-25$ (Thursday) (which site is denoted by the red dashed-dotted vertical line). Time plays, herein, the role of the control parameter. The empirical data are shown by the black erratic solid curve decorated by dots obtained for weekly horizon. The blue solid curve, consists of three parts (where the latter part is separated by the red solid one). The height of the first horizontal part equals $A_0=4.5273$ being a mean value of $k_{SZG}$ before the first transition time or the first critical threshold $t_{crit}= 164$[td]$\equiv ${2005-08-11 (Thursday)} (the location of this threshold is denoted by the first blue vertical dashed line). The second long-term part (of the order of one year), beginning at 
$t_{crit}$, was described by a power law function $A (t- t_{crit})^{1/z}+A_0$, where amplitude $A=2.50$ and the global dynamic exponent $z=2$. However, the early stage of semi-critical dynamics (of the order of one month denoted by the first and second blue dashed vertical lines as border ones) is driven by the canonical Lifshitz-Slyozov dynamic exponent $z=3$ and $A=5.20$. The third part or the right-hand side of the $\lambda $-peak is defined for 
$0<t-t_{\lambda }<\tau _R$ by the logarithmic relaxation function $-A_R\ln \left(\left(t-t_{\lambda }\right)/\tau _R\right)$, where $A_R=22$ and 
$\tau _R=96=480/5$[tw] (standard trading week consists of 5 trading days). This function is solution (\ref{rown:sollambdaR}) of macroscopic Eq. 
(\ref{rown:lambdaR}). The red impetuously increasing solid curve represents a logarithmic function 
$-A_L\ln \left(\left(t_{\lambda }-t\right)/\tau _L\right)$, for $0<t_{\lambda }-t\leq \tau _L$, where amplitude $A_L=14$ and $\tau _L=500=2500/5$[tw]. For both sides of the $\lambda $-peak the transition time $t_{\lambda }=110 =550/5$[tw]. The $k_{SZG}$ short-range cross over (of the order of one quarter) from the nucleation to condensation is placed within the surroundings of 2006-07 defined by the overlap of blue and red solid curves, where no sharp transition is observed.}
\label{figure:oba_zbocza_week}
\end{center}
\end{figure}

In Fig. \ref{figure:ZE_16} the snap-shot picture presents the empirical MST scale-invariant graph (placed in the lower row on the left-hand side of the figure) concerning sub-period ranging from 2004-10-28 to 2006-05-11, i.e. covering 400 trading days. The width of the scanning window is fixed at 400 trading days, for the entire time series duration, as it was found to be optimal (other widths equal 300, 350, 450, and 500 trading days were also used).
This graph is characterized by a power law  distribution of vertex degrees, with the exponent $\alpha =2.98$ (see the plot in the log-log scale placed in the upper row on the right-hand side of the figure\footnote{The temporal standard deviation, $\Delta \alpha (t)$, does not exceed of $10\%$ of the exponent $\alpha (t)$ for any time $t$.}). This figure clearly characterizes the situation typical for the sub-period named 'Equilibrium scale-invariant network' ranging from the left-hand boundary of Figs. \ref{figure:oba_zbocza} and \ref{figure:oba_zbocza_week}, i.e. from 2004-12-05 (Monday), to the first blue dashed vertical line located at 2005-08-11 (Thursday), that is for the MST network remaining in the equilibrium state -- this property is futher discussed in Sec. \ref{section:dbcond}. 
\begin{figure}
\begin{center}
\bigskip
\includegraphics[width=140mm,angle=0,clip]{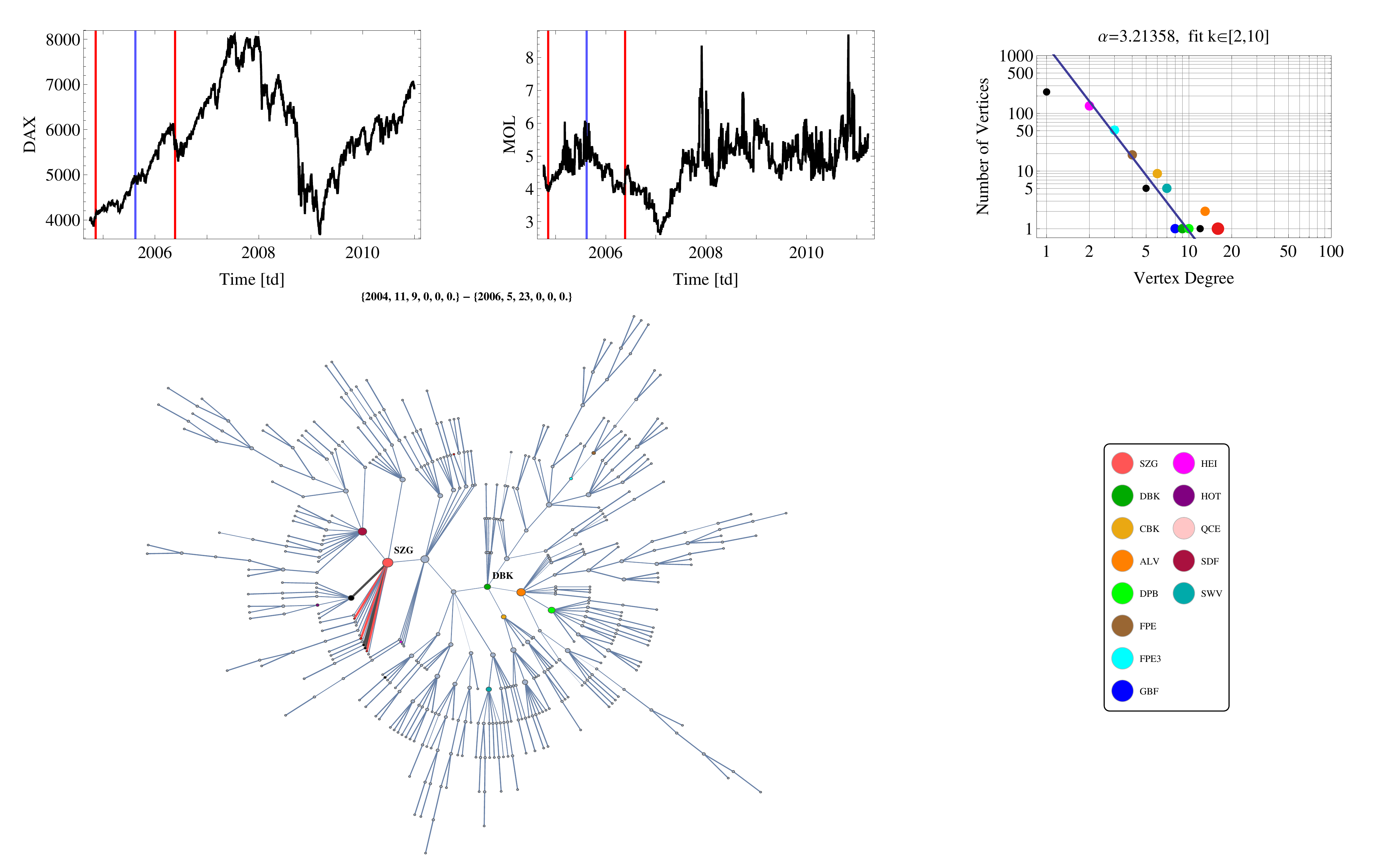}
\caption{Snap-shot picture of the FSE directly after the first two-day avalanche of edges attached by the SZG vertex at 2005-08-15 (Monday; the frame of the movie no. 24). The red edges denote the ones attached in the current step to the SZG vertex, while the black edge indicates the one which will be detached in the next step. Apparently, the SZG node became a leader shown by the red circle in the graph and also by the corresponding red circle slightly coming off the power-law in the plot placed in the upper row on its right-hand side. The description of other elements of this Figure is analogous to that given in Figs. \ref{figure:ZE_16} and \ref{figure:ZEP_402} as all of them are snap-shot pictures of the same movie.}
\label{figure:ZEP_24}
\end{center}
\end{figure}

Note that the size of the vertex in the graph is proportional to its degree. The circles of the same color, both in the graph and in the power law plot, represent the same company (their abbreviations are shown in the legend, while their corresponding names can be easily found in the internet). The  vertices which almost all the time occupy thirteen top positions of the rank are colored, while the remaining vertices are in grey (although some of them also occupy from time to time, but for very short time lag, a top position of the rank). Apparently, the two largest companies DBK (Deutsche Bank AG) and ALV (Allianz SE; green and orange circles, respectively placed in the centre of the graph) are direct neighbors for the period under consideration (i.e. for the `Equilibrium scale-invariant network'). That occurs, when the strongest mutual correlations are shown between the largest companies, which effectively, are capable of  balancing (or stabilizing) the entire stock market. 

Let's focus on the SZG company (very small red circle on the graph {in Fig. \ref{figure:ZE_16}}), which is now a marginal player since its degree hardly equals 3 (see also the red circle located in the power law plot in the log-log scale placed in the upper row on the right-hand side of the figure), but quickly (see Fig. \ref{figure:ZEP_24} for details) becomes a dominating vertex of the graph for about one-and-a-half year (see the sub-periods named `Nucleation' and `Condensation' in Figs. \ref{figure:oba_zbocza} and \ref{figure:oba_zbocza_week}). Indeed, we will systematically follow the `career' of this vertex by using characteristic snap-shot pictures produced by our empirical based simulation of the MST network evolution. The simulation was constructed from pictures prepared subsequently from empirical temporal daily (and, for self-consistency, also from some weekly) MSTs. To emphasize the analysis, each snap-shot picture is supplemented with the time-dependent plots of DAX and MOL (the upper row in each figure containing the MST graph).

\begin{figure}
\begin{center}
\bigskip
\includegraphics[width=140mm,angle=0,clip]{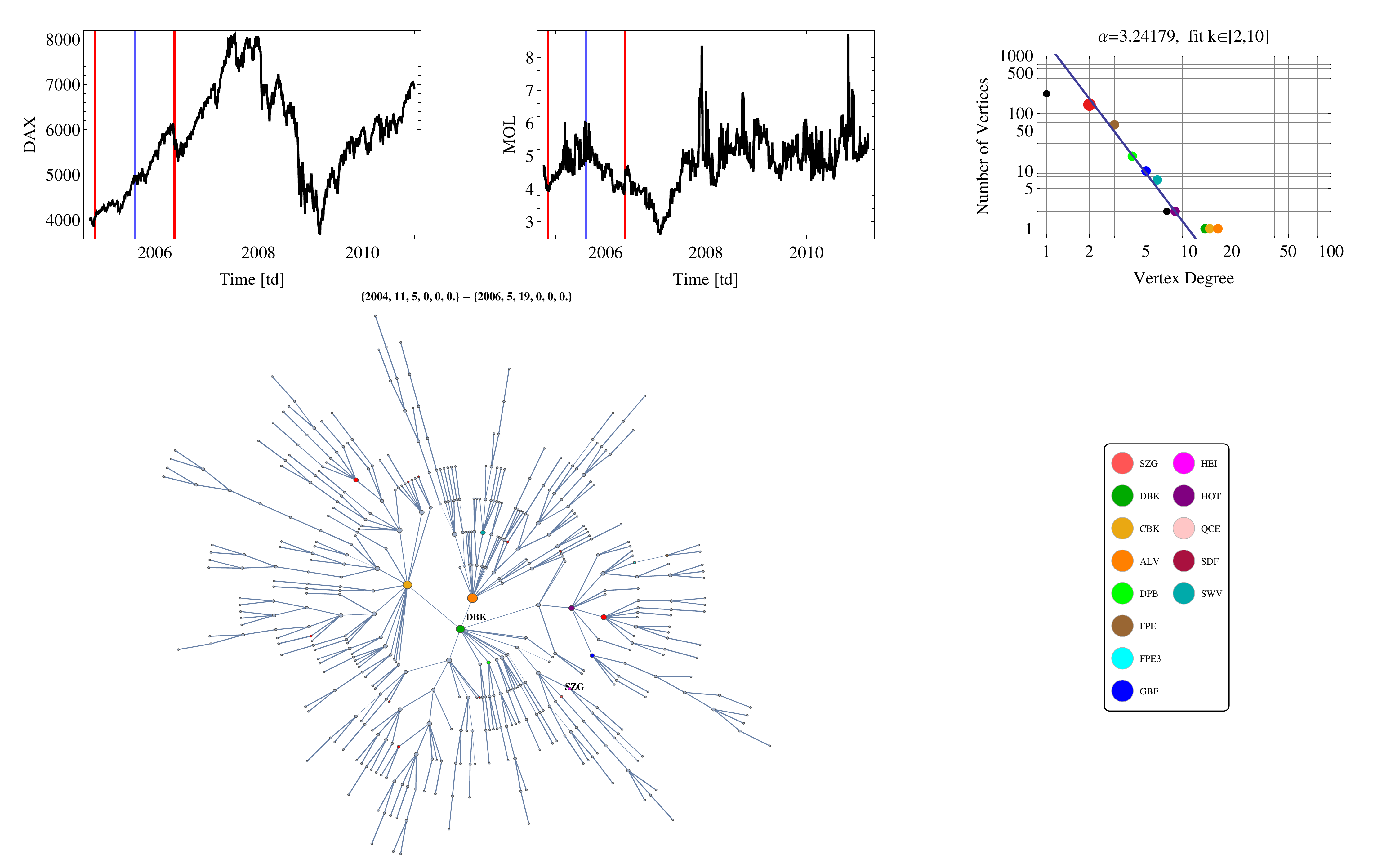}
\caption{Snap-shot picture of the FSE at 2005-08-11 (Thursday; the frame of the movie no. 22) i.e., one day before of the first day of the two-day avalanche of edges attached by the SZG vertex. The final stage of this avalanche is shown in Fig. \ref{figure:ZEP_24}. Apparently, the SZG company is a peripheral vertex having degree, which hardly equals 2 (see also the plot in the log-log scale placed in the upper row on its right-hand side) -- it becomes the richest one only after the avalanche ends (for comparison see Fig. \ref{figure:ZEP_24}).}
\label{figure:ZE_22}
\end{center}
\end{figure}

\begin{figure}
\begin{center}
\bigskip
\includegraphics[width=140mm,angle=0,clip]{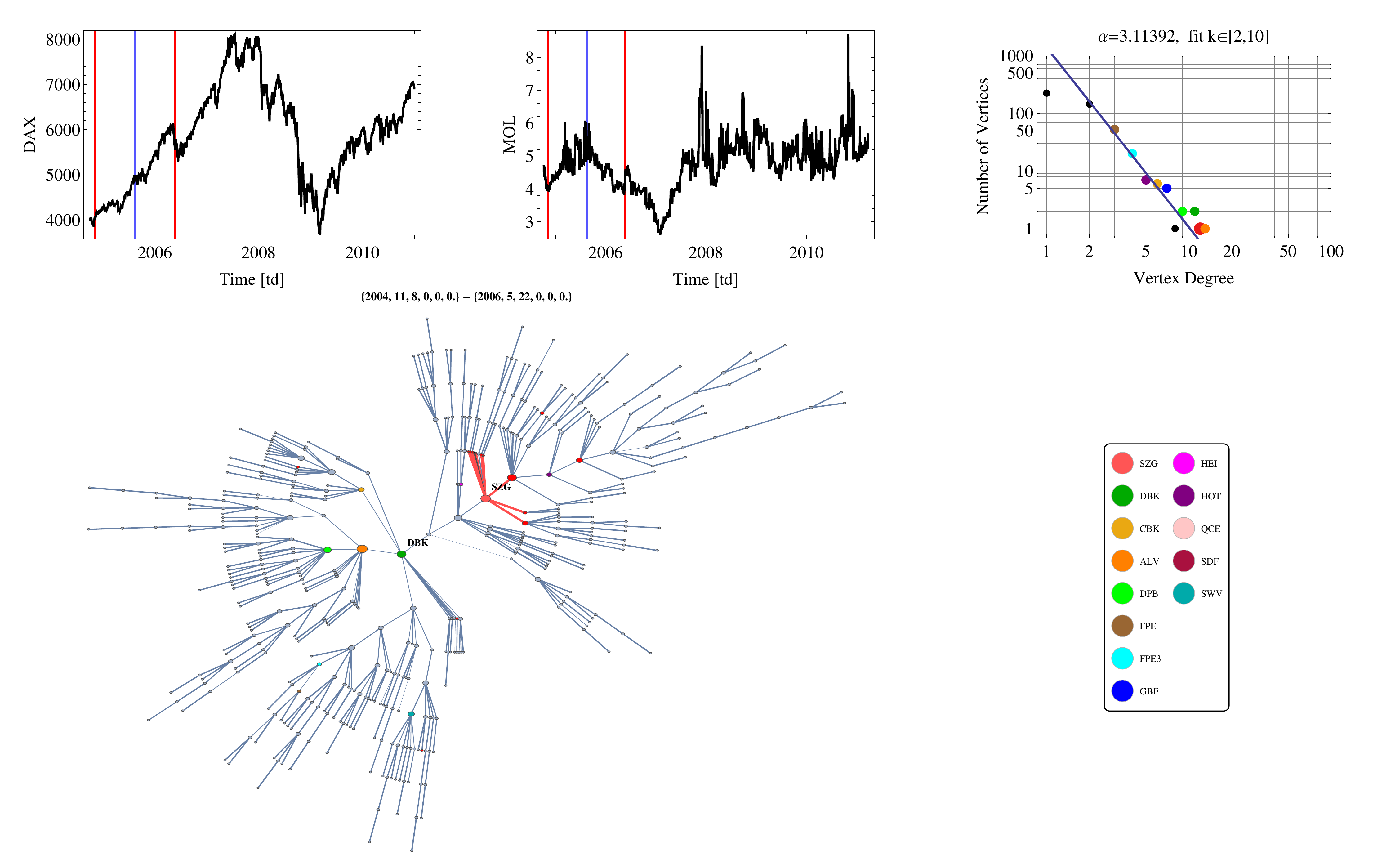}
\caption{Snap-shot picture of the FSE directly after the first day of the two-day avalanche of edges attached by the SZG vertex at 2005-08-12 (Friday; the frame of the movie no. 23). Although the SZG node increased its degree to 12, it is still not the richest one (it is a vice-leader at the moment). Only after the second day of the avalanche, it occupies already the leader position (see Fig. \ref{figure:ZEP_24} for details; further description of the Figure is analogous to that, e.g., of Fig. \ref{figure:ZEP_24}).}
\label{figure:ZE_23}
\end{center}
\end{figure}

The leader position was reached, for the first time, by SZG within the very narrow region, extended in Figs. 
\ref{figure:oba_zbocza} and \ref{figure:oba_zbocza_week}, between the first and second blue vertical dashed lines. This position was reached in two stages. The first stage, when SZG degree abruptly increased from 2 to 12 within one critical day from $t_{crit}\equiv $Thursday 2005-08-11 (cf. Fig. 
\ref{figure:ZE_22}) to Friday 2005-08-12 (cf. Fig. \ref{figure:ZE_23}) and the second stage from Friday 2005-08-12 to Monday 2005-08-15 (cf. Fig. 
\ref{figure:ZEP_24}) when its degree again increased but now from 12 to 16. This is sufficiently easily seen, since the edges in red denote the ones attached in the current step to the SZG vertex, while edge in black is going to be detached in the next step. Indeed, Thursday 2005-08-11 we can consider as the beginning of the 'Nucleation' sub-period, which breaks the time translation invariance. It is a beginning of the increase of the MST network order, being an analog of the phase-ordering \cite{DS0,HP}. 

During the evolution over the `Nucleation' and `Condensation' sub-periods (again see Figs. \ref{figure:oba_zbocza} and 
\ref{figure:oba_zbocza_week} for details), the SZG company still occupies the leader position, increasing its degree (up to some fluctuations). However, the way of this increase distinctly differs for both sub-periods. 

\begin{figure}
\begin{center}
\bigskip
\includegraphics[width=140mm,angle=0,clip]{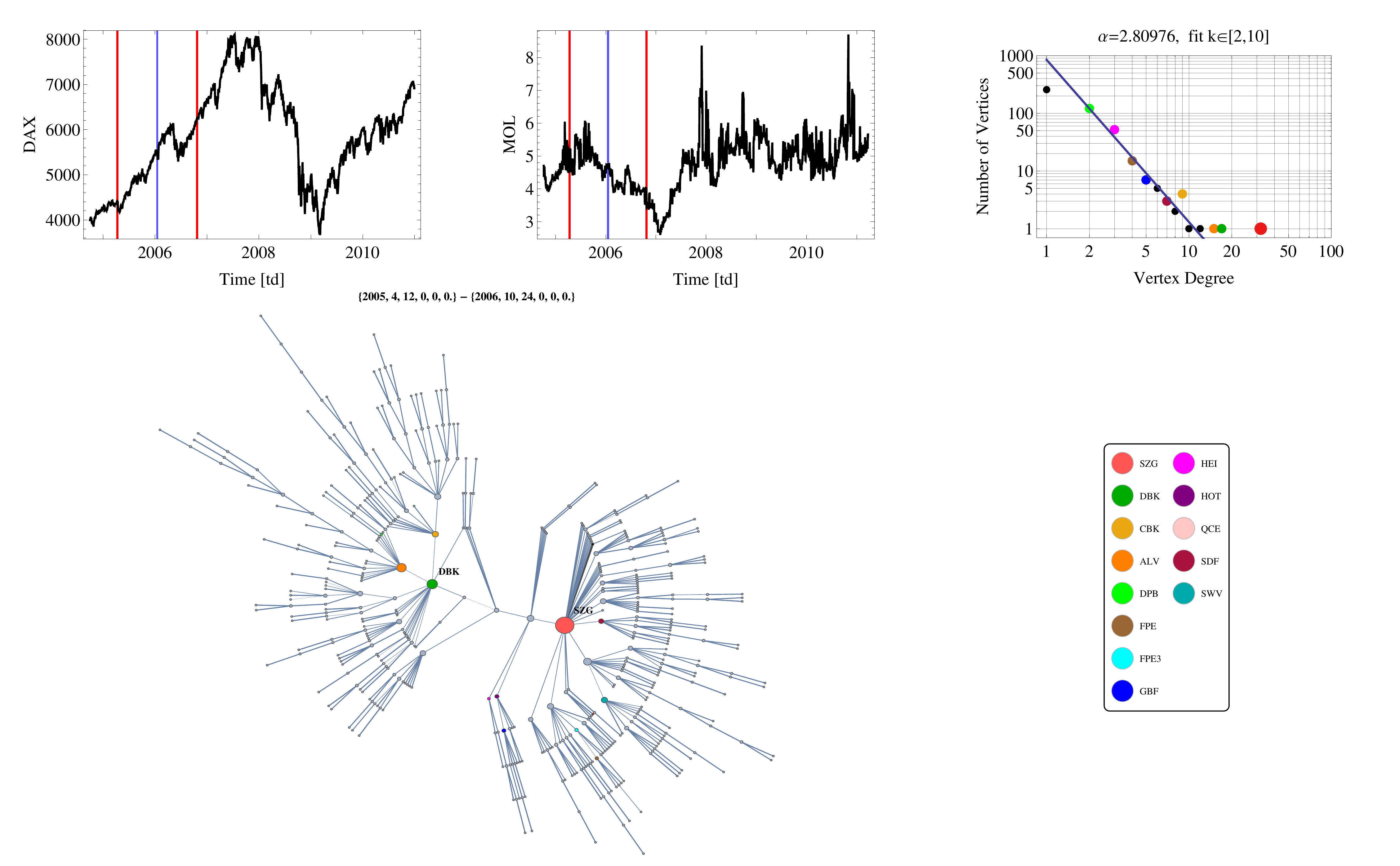}
\caption{Snap-shot picture of the empirical MST network (placed in the lower row on the left-hand side of the Figure) consisting of companies quoted on the FSE within the sub-period from 2005-04-12 to 2006-10-24 -- these boundary dates are denoted, in both plots placed in the upper row, by the red vertical straight lines. The center of this sub-period -- at January 16 (Monday), 2006 (the frame of the movie no. 134) -- is denoted by the blue vertical straight line. Please note, how much the SZG company is now coming off the power law -- here, its degree equals 32. Besides, DBK and ALV companies (occupying the second and third positions in the rank, respectively) are only slightly coming off this power law.}
\label{figure:ZE_134}
\end{center}
\end{figure}

For the `Nucleation' sub-period, the degree of the SZG node only slowly, although systematically (up to some fluctuations), increases -- except of the three days of the abrupt increase considered above. This kind of increase is determined by the structure of the MST, where very rich vertices are located (for this sub-period) very far from the leading SZG node. This is documented by the typical situation visualized in Fig. \ref{figure:ZE_134}, where the richest nodes (DBK and ALV) are located four and five `handshakes' from the SZG node. 

If we notice that the SZG node mainly attaches nodes from its second and third coordination zones, it becomes clear why the SZG degree increases so slowly. In particular, the situation visualized in Fig. \ref{figure:ZE_134} indicates that only a single vertex (coming from the second coordination zone) will become connected to the SZG node in the next time step and no vertex will be disconnected. This is easily seen, since the former node is red while the latter one is black. As previously, both here and in the entire work, the edges in red denote the ones attached in the current step to the SZG vertex, while edge in black is going to be detached in the next step.

\begin{figure}
\begin{center}
\bigskip
\includegraphics[width=140mm,angle=0,clip]{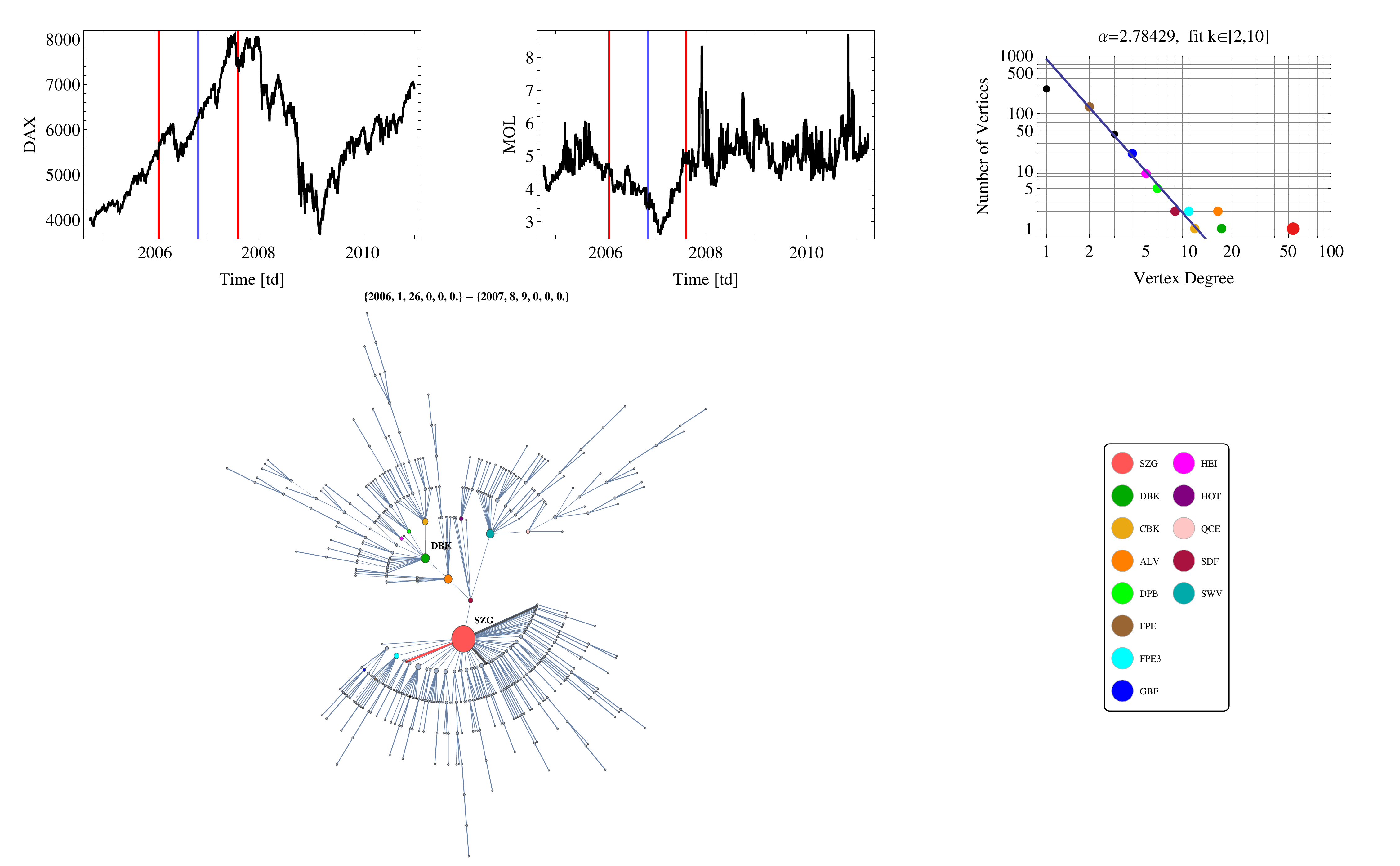}
\caption{Snap-shot picture of the empirical MST network (placed in the lower row on the left-hand side of the Figure) consisting of companies quoted on the FSE within the sub-period from 2006-01-26 to 2007-08-09 -- these boundary dates are denoted, in both plots placed in the upper row, by the red vertical straight lines. The center of this sub-period -- at November 2 (Thursday), 2006 (the frame of the movie no. 341) -- is denoted by the blue vertical straight line. Please note, how much the SZG company is now coming off the power law in comparison with earlier, analogous 
situations (for comparison see Fig. \ref{figure:ZE_134}) -- here, its degree equals 53. Besides, the DBK, ALV, and SWV companies (occupying the second and \emph{ex aequo} third and fourth positions in the rank, respectively) are only slightly coming off this power law. Furthermore, DBK and ALV are now `attracted' by SZG vertex to its second and third coordination layers (zones). They are now more closely located to the SZG vertex than earlier (for a detailed view please refer to Fig. 
\ref{figure:ZE_134}).}
\label{figure:ZEP_341}
\end{center}
\end{figure}

\begin{figure}
\begin{center}
\bigskip
\includegraphics[width=140mm,angle=0,clip]{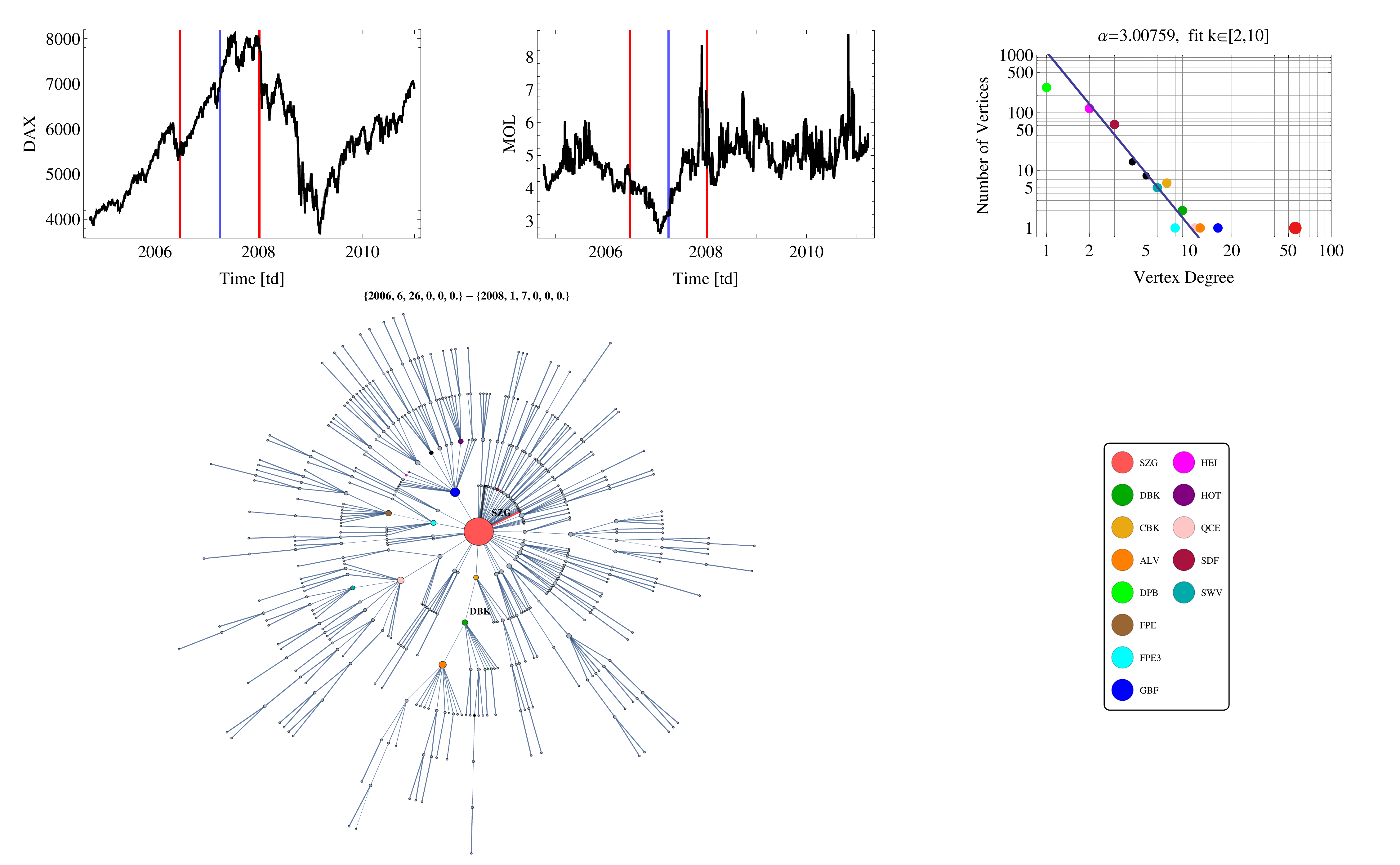}
\caption{Snap-shot picture of {the empirical MST network} (placed in the lower row on the left-hand side of the Figure) consisting of companies quoted on {the} FSE within the sub-period from 2006-06-26 to 2008-01-07 -- these boundary dates are denoted, in both plots placed in the upper row, by the red vertical straight lines. The center of this sub-period -- at April 2 (Monday), 2007 (the frame of the movie no. 448) -- is denoted by the blue vertical straight line. Notably, the SZG vertex returns towards the power law by reduction of its degree, however, it is still sufficiently large equaling 55. Apart from the SZG vertex, only the vice-leader (i.e., GBF company) is `attracted' to the first coordination zone of the SZG vertex. It is characteristic that after the MOL absolute minimum is passed (i.e., here about two months later), the ranking of companies drastically changed as the second position is now (although quite shortly) occupied by the one from {the} further position.}
\label{figure:ZEP_448}
\end{center}
\end{figure}

In Fig. \ref{figure:ZEP_341} we already present the situation characteristic for the `Condensation' sub-period. Apparently, four richest nodes (here DBK, ALV, SWV, CBK) are located, herein, in the second and third coordination zones of the SZG. This effect of the `attraction' is well visible at the MOL's absolute minimum and a bit later (cf. Figs. \ref{figure:ZEP_402} and  \ref{figure:ZEP_448}, respectively).
Furthermore, in the surroundings of the key date, i.e. 2007-01-25 (Thursday) denoted by the red dashed-dotted vertical line (e.g. in Figs. 
\ref{figure:entropy_deg_eff} -- \ref{figure:betweeness}), the vertices occupying the second position in the rank locate in the first or the second coordination layers.
Thus, the SZG company has at one's disposal a strongly increased number of edges. Presumably, this is the reason for the abrupt increase of its degree in the vicinity of $t_{\lambda }$, well seen in Figs. \ref{figure:oba_zbocza} and \ref{figure:kondensat_day}. 
\begin{figure}
\begin{center}
\bigskip
\includegraphics[width=170mm,angle=0,clip]{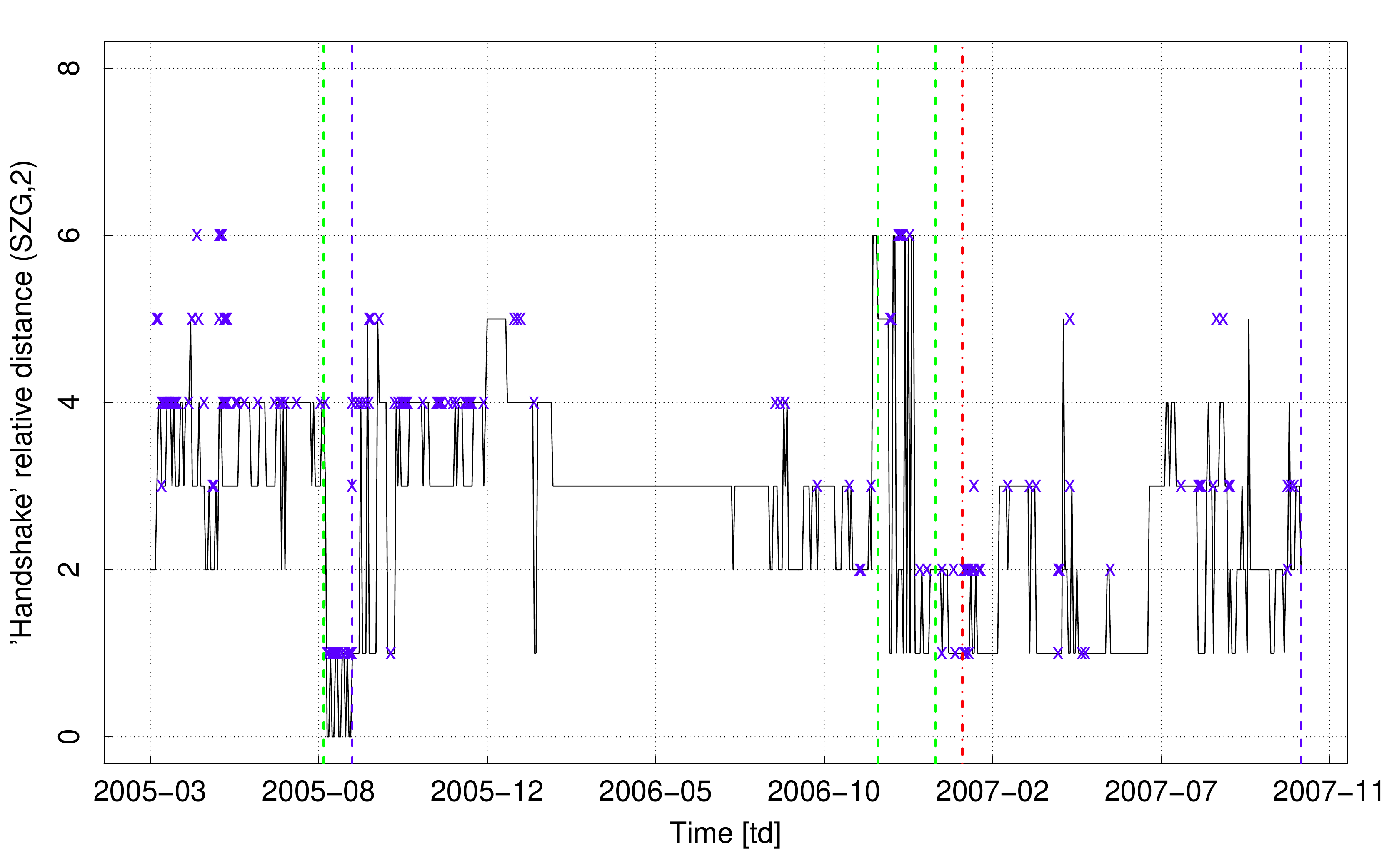}
\caption{The plot of the temporal relative distance `handshake' (THSD; presented by the erratic (jumping) curve) between SZG vertex and the temporary (first) vice-leader one (denoted by number $2$ in the description of the vertical axis). The erratic (jumping) curve connects the current values of the THSD, while crosses present the THSD of the twin vertex (i.e., the second vice-leader having the same degree but whose temporal `handshake' relative distance is not shorter than that of the first vice-leader). Apparently, in the surroundings of the key {critical} date, i.e. 2007-01-25 (Thursday) denoted by the red dashed-dotted vertical line, the vertices occupying the second position in the rank are located in the first or the second coordination layers. Thus, the SZG company has at one's disposal a strongly increased number of edges. This is the reason of the abrupt increase of its degree (within the range whose left and right boundary are denoted by the green vertical dashed lines located next to the key date -- see also Figs. \ref{figure:oba_zbocza}, 
\ref{figure:oba_zbocza_week}, \ref{figure:kondensat_day}, and \ref{figure:kondensat} for details), also supported by the analogous behavior displayed by vertices occupying the third position in the rank (see Fig. \ref{figure:SZG_k3_correct} for details). Notably, the first pair of dashed vertical lines defines the beginning of the nucleation phase -- see also Figs. \ref{figure:oba_zbocza}, \ref{figure:oba_zbocza_week}, \ref{figure:kondensat_day}, and 
\ref{figure:kondensat} for details. {There, the vanishing value of the THSD means that the SZG company occupies itself the second position in the rank. Notably, only for the sub-period limited by the two blue dashed vertical lines, the SZG company occupies the leader position of the rank.}}
\label{figure:SZG_k2_correct}
\end{center}
\end{figure}
\begin{figure}
\begin{center}
\bigskip
\includegraphics[width=170mm,angle=0,clip]{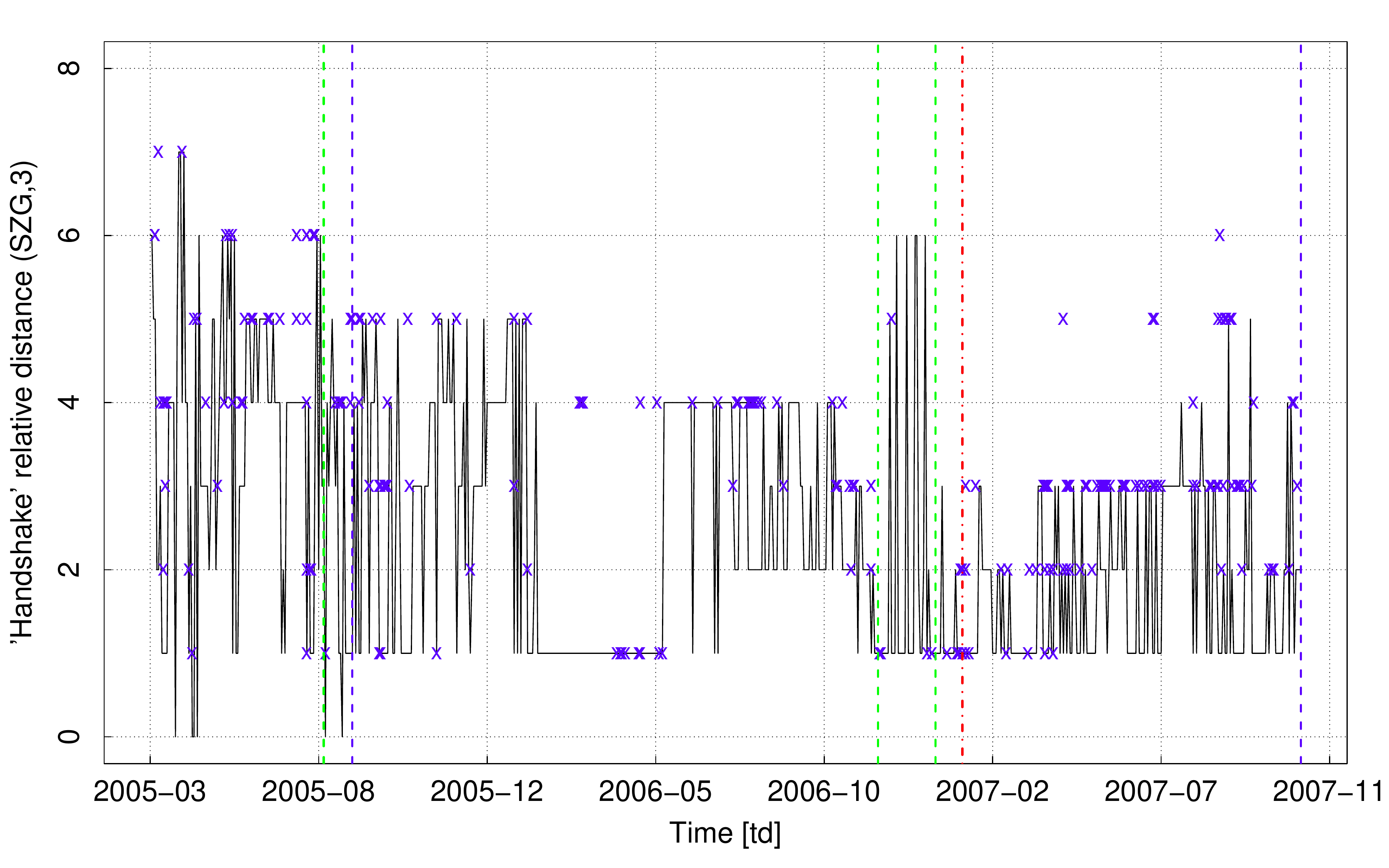}
\caption{The plot of the temporal relative distance `handshake' (THSD; presented by the erratic (jumping) curve) between SZG vertex and the one currently occupying the third position in the rank (denoted by number $3$ in the description of the vertical axis). The crosses represent another vertex occupying \emph{ex aequo} the third position in the rank but whose THSD from the SZG vertex is not shorter than the former despite having the same degree. Apparently, in the surroundings of the key date, i.e. 2007-01-25 (Thursday) denoted by the red dashed-dotted vertical line, the vertices occupying the third position in the rank are located in the first, second or third coordination layers. The pair of green vertical dashed lines located next to the key critical date and the first pair of the dashed lines are already defined in Fig. \ref{figure:SZG_k2_correct}. There, the vanishing value of the THSD means that the SZG company occupies itself the third position in the rank. Notably, only for the sub-period limited by two blue dashed vertical lines, the SZG company occupies the leader position of the rank.}
\label{figure:SZG_k3_correct}
\end{center}
\end{figure}

For prolonged time (when network passed $t_{\lambda }$), `repulsion' dominates `attraction' (cf. Fig. \ref{figure:ZEP_558} for details) and the MST network decouples into several locally centralized clusters (sectors) which leads to disintegration of the condensate.

\section{Kinetic equations of the market plankton}\label{section:keqpoor}

In this section we verify, based on the empirical evidence provided above, a usefulness of discrete kinetic equations, and hence power law distributions of vertex degrees, for study of the MST network and sub-network in equilibrium. For instance, we extract rules of edges' connection and disconnection for poor vertices, or market plankton, and study some of their most significant consequences. By the term `poor vertices' we define vertices whose degree distribution is a power law. We observed that during many-year evolution of the MST network, the power law distributions obey only vertices having degrees smaller than 12 -- there are a few rich vertices having larger degrees, which comes off the power law. 
In the following, we consider the properties of such a phase, denoted in Figs. \ref{figure:oba_zbocza}, \ref{figure:oba_zbocza_week}, 
\ref{figure:kondensat_day}, and  \ref{figure:kondensat} as the `Equilibrium scale-free network' and the influence of this phase on the other phases.

\subsection{Bulk versions of the kinetic equation}\label{section:Bulkkin}

We propose a discrete kinetic equation of the Markovian type for the probability, 
$P(k,t)$, of finding at time $t$ a vertex of complex network, which has degree equals $k$, where $k=2\ldots , n-2$, defining the bulk version of the kinetic equation (here $n$ is a number of vertices of the MST network). That is, the boundary cases defined by $k=1$ and $k=n-1$ are considered in section \ref{section:Border}. Hence,
\begin{eqnarray}
P(k,t+1)&=&\sum_{l=1}^{k-1}p(k,t+1|k-l,t) P(k-l,t) \nonumber \\ 
&+&\sum_{l=1}^{n-1-k}p(k,t+1|k+l,t) P(k+l,t) \nonumber \\ 
&+&p(k,t+1|k,t) P(k,t) \Leftrightarrow \nonumber \\
P(k,t+1)&-&P(k,t)= -\left[J_{out}(k;t,t+1) - J_{in}(k;t,t+1)\right],
\label{rown:kemt}
\end{eqnarray}
where $p(k,t+1|k\pm l,t)$ is a single-step gain transition probability from degree $k\pm l$ to $k$, and $p(k,t+1|k,t)$ is a single-step survival probability of a vertex having degree $k$. Here we are dealing with a ladder model where jumps over several rungs are possible. 

The continuity equation in Eq. (\ref{rown:kemt}), was derived by substituting a single-step survival probability taken from the normalization condition containing the corresponding loss transition probabilities,
\begin{eqnarray}
p(k,t+1|k,t)=1-\sum_{l=1}^{n-1-k}p(k+l,t+1|k,t)-\sum_{l=1}^{k-1}p(k-l,t+1|k,t),
\label{rown:norm}
\end{eqnarray}
where last two terms describe all possible ways of abandonment of a given vertex having degree equaling $k$.

The macroscopic (effective, global) single-step currents present in the continuity equation, flowing out, $J_{out}$, of vertex degree $k$ and into this degree, $J_{in}$, are defined as follows
\begin{eqnarray}
J_{out}(k;t,t+1) = \sum_{l=1}^{n-1-k}j(k+l,t+1;k,t) 
\end{eqnarray}
and
\begin{eqnarray}
J_{in}(k;t,t+1) = \sum_{l=1}^{k-1}j(k,t+1;k-l,t),
\label{rown:currents}
\end{eqnarray}
respectively. The microscopic (effective, local) single-step currents from $k$ to $k+l$ 
\begin{eqnarray}
j(k+l,t+1;k,t)&=&p(k+l,t+1|k,t)P(k,t) \nonumber \\
&-&p(k,t+1|k+l,t)P(k+l,t),\; l=1, \ldots , n-1-k, \nonumber \\
\label{rown:currentout}
\end{eqnarray}
and from $k-l$ to $k$
\begin{eqnarray}
j(k,t+1;k-l,t)&=&p(k,t+1|k-l,t) P(k-l,t) \nonumber \\
&-&p(k-l,t+1|k,t) P(k,t),\; l=1, \ldots , k-1,
\label{rown:currentin}
\end{eqnarray}
constitute an avalanche of microscopic currents (or cascading microscopic flows), which if positively oriented, are of loss and gain types, respectively. Notably, the continuity equation is associated here with the conservation of the total number of vertex degrees, which equals (for arbitrary MST network) $2(n-1)$, where $n$ is the total number of vertices. Detailed balance conditions, which we explore in Sec. \ref{section:dbcond}, have to be consistent with equations in (\ref{rown:kemt}). 

\subsection{Boundary versions of the kinetic equation}\label{section:Border} 

The boundary versions of the kinetic equation only concerns boundary vertex degrees $k=1$ and $k=n-1$. These versions are the particular case of the first equation in Eq. (\ref{rown:kemt}) and formally identical to the continuity (second) equation where, however, a restricted definitions of currents were used. Namely,
\begin{eqnarray}
J_{out}(k;t,t+1) = \sum_{l=1}^{n-2}j(k+l,t+1;k,t),\; \mbox{for $k=1$}, \nonumber \\
J_{in}(k;t,t+1) = \sum_{l=1}^{n-2}j(k,t+1;k-l,t),\; \mbox{for $k=n-1$},
\label{rown:bcurrents}
\end{eqnarray}
where
\begin{eqnarray}
j(k+l,t+1;k,t)&=&p(k+l,t+1|k,t) P(k,t) \nonumber \\
&-&p(k,t+1|k+l,t) P(k+l,t),\; \mbox{for $k=1$},\nonumber \\
j(k,t+1;k-l,t)&=&p(k,t+1|k-l,t)P(k-l,t) \nonumber \\
&-&p(k-l,t+1|k,t)P(k,t),\; \mbox{for $k=n-1$}. 
\label{rown:bcurrentsd}
\end{eqnarray}
Obviously, the restriction is caused only by the bottom and top of ladder's rungs. In principle, in Sec. 
\ref{section:dbcond}, the detailed balance conditions also use the boundary vertex degrees. 

\subsection{Detailed balance conditions}\label{section:dbcond}

In the first stage, we restrict our considerations to vertex degrees smaller than $k=12$ as, approximately, for range $2\leq k < 12$, the distribution of vertex degrees is satisfactorily represented by the power law function (for the illustration see Figs. \ref{figure:ZE_16} -- \ref{figure:ZEP_558}, 
\ref{figure:ZEP_24} -- \ref{figure:ZEP_448}; the boundary vertices, having degrees $k=1$ or $k=n-1$, are much less significant). Apparently, our empirical results give, at the range of $k$ considered, such a robust power law behavior, that we can assume sub-network, consisting of the corresponding vertices, as being at least in partial equilibrium\footnote{Then, $\ln k$ can play the role of energy, i.e. 
$\varepsilon (k) \equiv \ln k$ and exponent $\alpha -1$ plays the role of an inverse temperature $\beta $, i.e $\beta =\alpha -1$. Hence, the Boltzmann-Gibbs probability distribution $\tilde{P}(\varepsilon )= \exp(-\beta \varepsilon )/Z$, where partition function $Z=\sum_{\varepsilon =0}^{\ln (n-1)}\exp(-\beta \varepsilon )$. In this derivation, we use the relation $\tilde{P}(\varepsilon )=P(k(\varepsilon )) \frac{dk(\varepsilon )}{d\varepsilon }$, where 
$k(\varepsilon )= \exp (\varepsilon )$. Having the partition function, we can develop the statistical thermodynamics of the MST network with fixed number of vertices analogous to that developed by Albert-Barab\'asi \cite{RAALB} for the analysis of Bose-Einstein condensation in the growing network, if additionally, the proper condition of quantum-mechanical undistinguishability of edges was used. Furthermore, the Landau criterion for superfluidity \cite{IChS} can be found by changing the variable $k$ to $n-1-k$. Nevertheless, our approach developed in this work is the alternative one, easier for empirical verification.}. Hence, (to good approximation) detailed balance conditions should be valid by putting all microscopic currents as vanishing, i.e. those from $k$ to $k+l$
\begin{eqnarray}
j(l;k)=0 \Leftrightarrow p(l|k)=p(-l|k+l)\frac{P(k+l)}{P(k)},\;l=1, \ldots , n-1-k, 
\label{rown:currentout0}
\end{eqnarray}
and similarly, the ones from $k-l$ to $k$ 
\begin{eqnarray}
j(l;k-l)=0 \Leftrightarrow p(l|k-l)=p(-l|k) \frac{P(k)}{P(k-l)},\; l=1, \ldots , k-1,
\label{rown:currentin0}
\end{eqnarray}
where $j(l;k)$ and $j(l;k-l)$ are stationary versions of the respective time-dependent currents $j(k+l,t+1;k,t)$ and 
$j(k,t+1;k-l,t)$. The analogous correspondence concerns the transition rates, that is $p(l|k)$, 
$p(-l|k+l)$, $p(l|k-l) $, and $p(-l|k)$ which are stationary versions of the time-dependent transition rates 
$p(k+l,t+1|k,t)$, $p(k,t+1|k+l,t)$, $p(k,t+1|k-l,t)$, and $p(k-l,t+1|k,t)$, respectively. The transformation from dynamic to static quantities is also made for the corresponding degree distributions.

\begin{figure}
\begin{center}
\bigskip
\includegraphics[width=120mm,angle=0,clip]{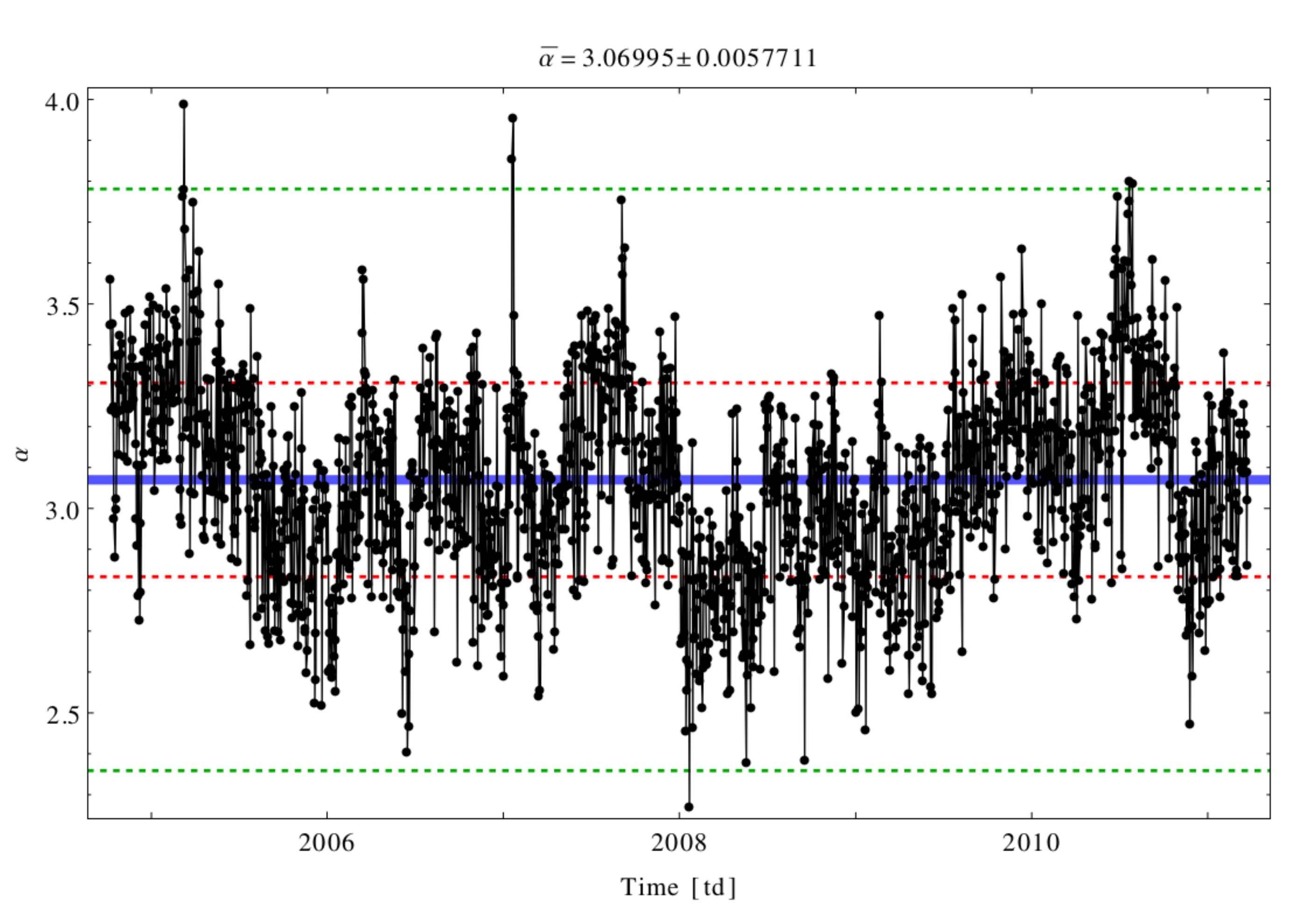}
\caption{Exponent $\alpha $, vs. time $t$. Its mean value $\bar{\alpha }=3.07$ is marked by the thick blue horizontal straight line. Indeed, this mean value obtained for the entire period of the MST network evolution was used in both expressions in Eq. (\ref{rown:rdegreed}). The positive and negative dispersions, $\pm \sigma $, of the data were denoted by the thin red dashed horizontal straight lines, while $\pm 3\; \sigma $ by the thick green dashed horizontal lines.}
\label{figure:mean_alpha}
\end{center}
\end{figure}

It is significant for the study of equilibrium properties of the MST network, that the ratios of the degree distributions, given in Eqs. (\ref{rown:currentout0}) and (\ref{rown:currentin0}), can be obtained as the corresponding ratios of power laws
\begin{eqnarray}
\frac{P(k+l)}{P(k)}=\left(1+\frac{l}{k}\right)^{-\bar{\alpha }},\; l=1, \ldots , n-1-k, \nonumber \\
\frac{P(k)}{P(k-l)}=\left(1-\frac{l}{k}\right)^{\bar{\alpha }},\; l=1, \ldots , k-1,
\label{rown:rdegreed}
\end{eqnarray}
where exponent $\bar{\alpha }$ was taken directly from empirical power laws as an average over all temporal exponents $\alpha (t)$ (cf. Fig. \ref{figure:mean_alpha}). Obviously, this is a crude empirical approach -- more accurate would be the calculation of $\bar{\alpha }$ for each phase of the MST network separately.

In the second stage we derive transition probabilities for disconnections of $l$ edges present in Eqs. (\ref{rown:currentout0}) and 
(\ref{rown:currentin0}). Thus, by using the pure binomial strategy, 
\begin{eqnarray}
p(-l|k+l)&=&
\left(
\begin{array}{c}
k+l \\
l
\end{array}
\right)
b(-1|k+l)^l (1-b(-1|k+l))^k, \nonumber \\
l&=&1, \ldots , n-1-k,  \nonumber \\
p(-l|k)&=&
\left(
\begin{array}{c}
k \\
l
\end{array}
\right)
b(-1|k)^l (1-b(-1|k))^{k-l},\; l=1, \ldots , k-1.
\label{rown:trdegreed}
\end{eqnarray}
Furthermore, within this strategy we have
\begin{eqnarray}
p(l|k)=
\left(
\begin{array}{c}
n-1-k \\
l
\end{array}
\right)
b(1|k)^l (1-b(1|k))^{n-1-k-l},\; l=1, \ldots , n-1-k.
\label{rown:trdegreedX}
\end{eqnarray}

The results obtained above are fairly universal, as they are obtained without using any detailed balance condition, only by assuming probabilities 
$b(-1|k+l)$ and $b(-1|k)$ for disconnection of a given single edge from vertex having degree equal $k+l$ or $k$, respectively, as known basic quantities. Obviously, having $b(-1|k)$ for $2\leq k\leq n-1$, we can already calculate shifted quantity $b(-1|k+l)$ for $2\leq k+l\leq n-1$. In fact, our basic quantities are the corresponding $b(\mp 1|k)$ ones, where $b(1|k),\ k=1,\ldots , n-2$, is the transition probability for the  connection of a given single edge to the vertex having degree equaling $k$. 

\begin{figure}
\begin{center}
\bigskip
\includegraphics[width=130mm,angle=0,clip]{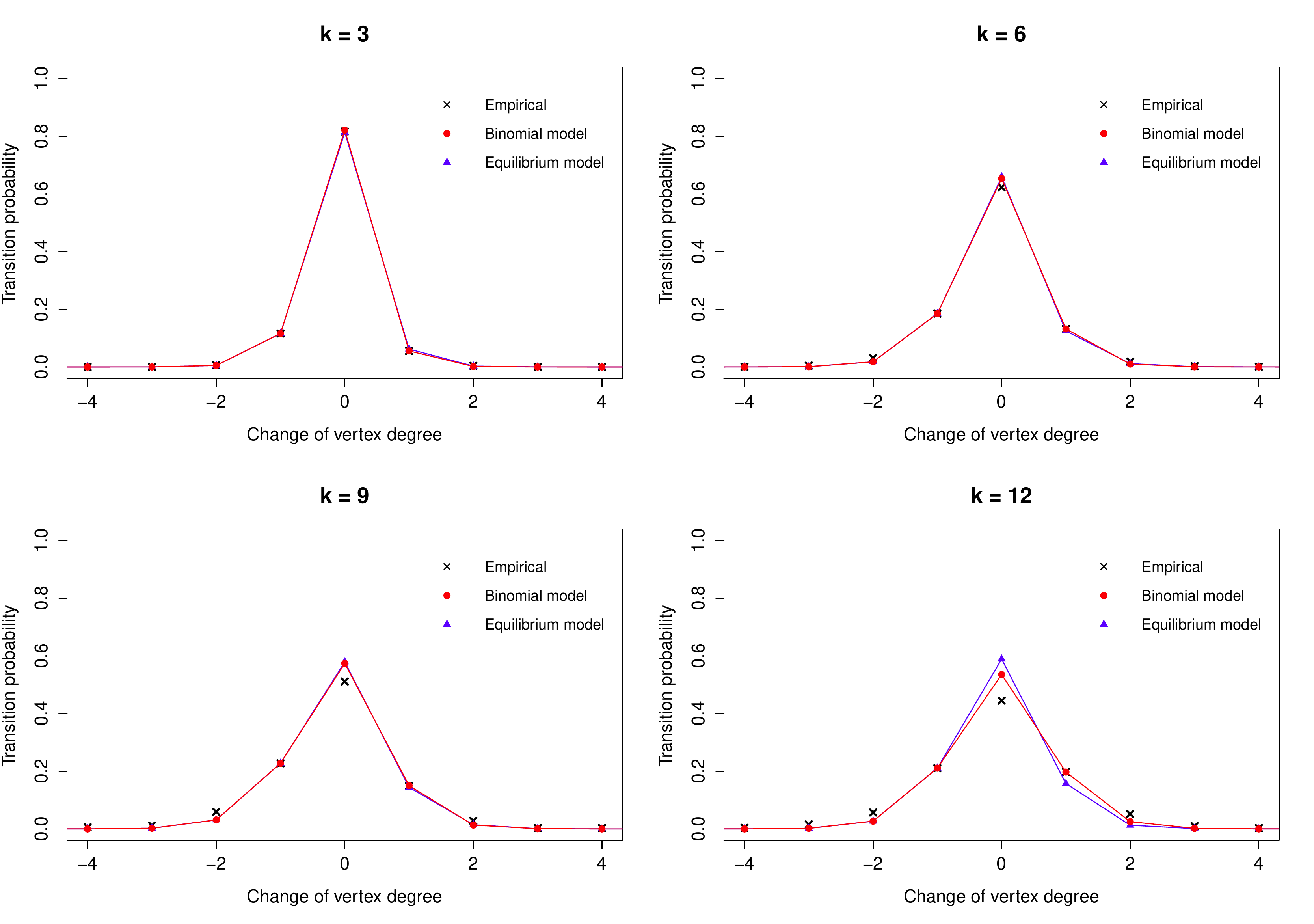}
\caption{The transition probability, $p(\pm l|k)$, vs. change of the vertex degree, $\pm l,\; l=0, 1, 2, \ldots $, for the MST network for four typical values of vertex degree $k=3,6,9,$ and $12$, that is, for degrees belonging only to the set of poor vertices. The crosses denote empirical quantities, while solid curves represent quite well fitting theoretical  predictions (described in details in the main text).}
\label{figure:deg_ch_prob-1}
\end{center}
\end{figure}

\begin{figure}
\begin{center}
\bigskip
\includegraphics[width=120mm,angle=0,clip]{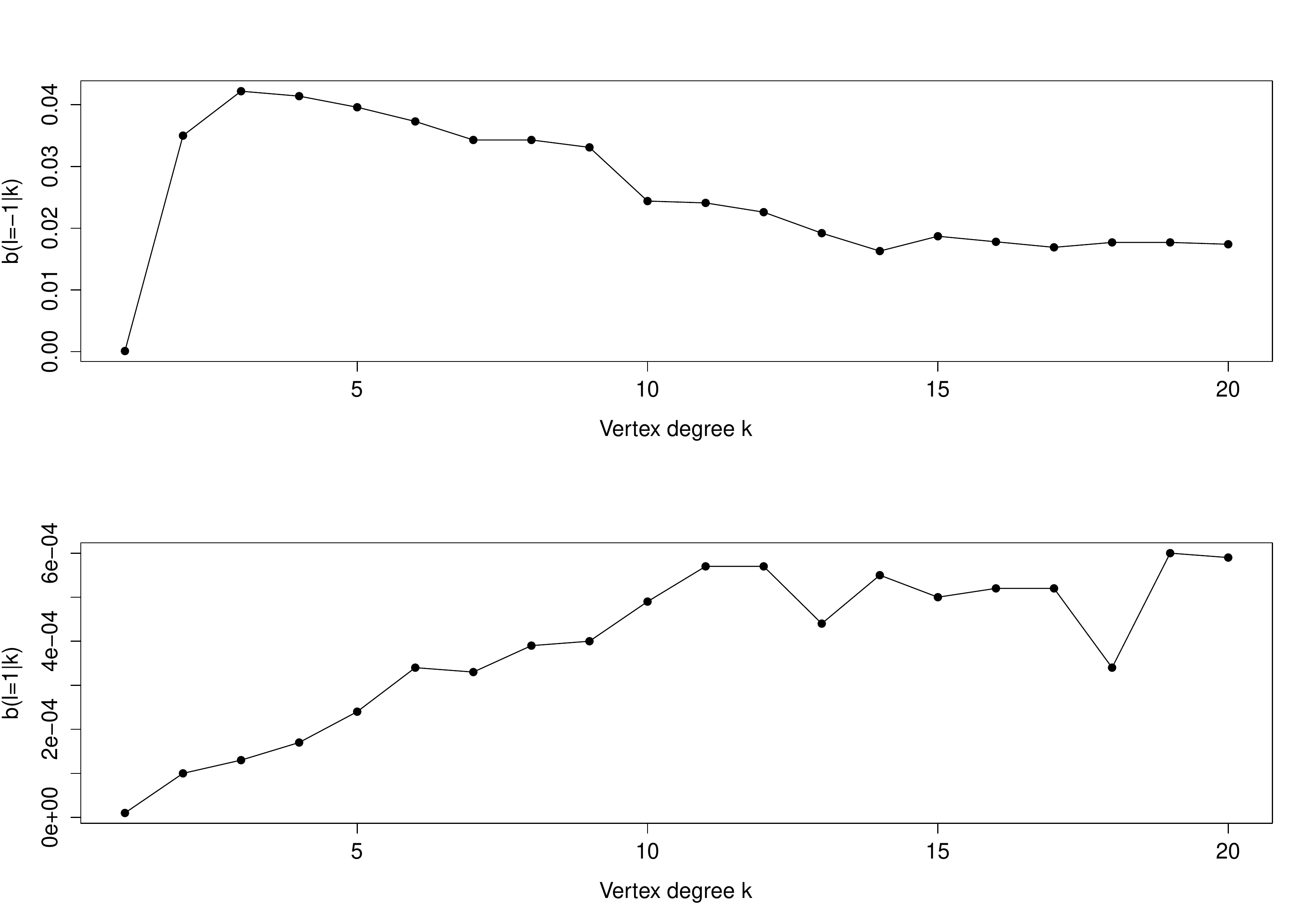}
\caption{The basic transition probabilities, $b(\mp 1|k)$, of disconnection and connection of a given single edge from or to the vertex of degree $k$ vs. $k$ (for $1\leq k\leq 20$) presented, respectively, in the upper and lower plots. For larger $k$, the statistics was, unfortunately, insufficient to construct these transition probabilities.}
\label{figure:fit-1}
\end{center}
\end{figure}

To derive the quantities necessary to calculate the transition probabilities for the connection of $l$ edges, $p(l|k)$ and $p(l|k-l)$, present in Eqs. 
(\ref{rown:currentout0}) and (\ref{rown:currentin0}), respectively, we need only transition probability $b(-1|k)$ for $2\leq k\leq n-1$. This quantity is directly obtained for the negative change of vertex degree $-l <0$, from the fit of $p(-l|k)$, given by the second expression in Eq. (\ref{rown:trdegreed}) -- all theoretical predictions are shown in Fig. \ref{figure:deg_ch_prob-1} by solid curves -- to their empirical counterparts, where all empirical data are shown in Fig. \ref{figure:deg_ch_prob-1} by crosses. The quantity $b(-1|k)$ calculated on this basis was plotted in Fig. \ref{figure:fit-1} using dots joined with a black line.

Substituting the Eqs. (\ref{rown:rdegreed}) and (\ref{rown:trdegreed}) into Eqs. (\ref{rown:currentout0}) and 
(\ref{rown:currentin0}), respectively, we finally find 
\begin{eqnarray}
p(l|k)&=&
\left(
\begin{array}{c}
k+l \\
l
\end{array}
\right)
b(-1|k+l)^l (1-b(-1|k+l))^k \left(1+\frac{l}{k}\right)^{-\bar{\alpha }}, \nonumber \\
l&=&1, \ldots , n-1-k,  \nonumber \\
p(l|k-l)&=&
\left(
\begin{array}{c}
k \\
l
\end{array}
\right)
b(-1|k)^l (1-b(-1|k))^{k-l}\left(1-\frac{l}{k}\right)^{\bar{\alpha }},\; l=1, \ldots , k-1. 
\label{rown:finaltrans}
\end{eqnarray}
These results we obtained with support of the detailed balance conditions, therefore, they are valid only for the network or sub-network in statistical equilibrium, which can be directly verified by comparison with the corresponding empirical data (shown in Fig. \ref{figure:deg_ch_prob-1} by crosses). 

It should be emphasized, that independently derived from our empirical data is not only the (above considered) single-step transition probability of jump down on the ladder, $p(-l|k)$, but also that of an upward jump, $p(l|k)$. The complete set of their empirical forms are shown in Fig. \ref{figure:deg_ch_prob-1} by crosses. Substituting these probabilities into the normalization condition (\ref{rown:norm}) we obtained the empirical single-step survival probability, 
\begin{eqnarray}
p(0|k)=1-\sum_{l=1}^{n-1-k}p(l|k)-\sum_{l=1}^{k-1}p(-l|k),
\label{rown:normhk}
\end{eqnarray}
presented in all plots in Fig. \ref{figure:deg_ch_prob-1} by the central crosses. Obviously, Eq. (\ref{rown:normhk}) is also valid for the theoretical single-step survival probability.

As $b(-1|k)$ and $\bar{\alpha }$ (plotted in Figs. \ref{figure:mean_alpha} and \ref{figure:fit-1}, respectively) were already earlier obtained from empirical data, both expressions in (\ref{rown:finaltrans}) have no free parameters. Hence, such a good agreement between their predictions (blue small circles connected by blue solid line) and empirical data (crosses), which is well seen in Fig. \ref{figure:deg_ch_prob-1} for vertex degree 
$2\leq k < 12$, 
means\footnote{The agreement would be improved if we prepare the empirical data for each MST network phase separately.} that poor vertices (defined by $2\leq k < 12$) are, in fact, in equilibrium forming a kind of a background for few rich vertices (having huge degrees equal or greater than $12$). These rich vertices are markedly off the power laws, as shown, for instance, in Figs. \ref{figure:ZEP_558}, \ref{figure:ZE_134} -- \ref{figure:ZEP_448}.

For completeness, we also calculated $b(1|k),\; k=1,\ldots , n-1$, analogously to the calculation of $b(-1|k)$. That is, we well fit the prediction of formula 
(\ref{rown:trdegreedX}) (blue small triangles connected by blue segments of solid line located on the positive parts of plots in Fig. 
\ref{figure:deg_ch_prob-1}) to the corresponding empirical data (crosses in the same figure) deriving, as a result, the proper values of $b(1|k)$ (for the visualization, see Fig. \ref{figure:fit-1}). The central blue triangle was also obtained (see Fig. \ref{figure:deg_ch_prob-1}) from the normalization condition (\ref{rown:normhk}). Hence, we verified both the binomial strategy, and the equilibrium hypothesis. This verification would be more convincing had we prepared the empirical data for each phase of the MST network separately.

As it is seen from Fig. \ref{figure:deg_ch_prob-1}, for each $k$, there exists an upper cut off for the positive change of the vertex degree, $l_{max}(k)$,  which makes the transition prabability, $p\left(l_{max}(k)|k\right)$, vanishing (to a good approximation). The analogous feature possesses a lower cut off, $l_{min}(k)$, which makes transition probability, $p\left(-l_{min}(k)|k\right)$, vanishing. Results presented in Fig. \ref{figure:deg_ch_prob-1} suggest that both $l_{max}(k)$ and $l_{min}(k)$ do not exceed 5 (to a good approximation). Larger $l_{max}(k)$ (but not exceeded 10) was observed only for few richest nodes. These restrictions make numerical and analytical calculations significantly easier.

To conclude this section, we can say that for poor vertices, the probability of edges' connection is much smaller than the probability of their disconnection (cf. comparison in Fig. \ref{figure:fit-1}) although, the former is a monotonically increasing function of the vertex degree. As the total number of edges is constant, it causes that only richest nodes are more and more richer.  
 
Furthermore, by using a rough analogy, we can say that we deal with two different kinds of fluid, which appears to be formally similar to those introduced both by London and Tich{\'y} \cite{KHuang} (and refs. therein) in their two-fluid models of Helium II (that is, below $\lambda $-transition in He$^4$). The sub-network of the poor vertices corresponds here to the normal fluid component, while the richest sub-network to the superfluid. In our approach both 
sub-networks are considered as coupled and, in general, disconnected ones. In the next Section we mainly discuss the network below the dynamic 
$\lambda $-transition \cite{MRRK} (and refs. therein), which we found herein (see Figs. \ref{figure:oba_zbocza}, \ref{figure:oba_zbocza_week}, 
\ref{figure:kondensat_day}, \ref{figure:kondensat}). However, the question whether the mapping of the MST network onto the bosonic lattice gas 
\cite{MRRK,BLG} is possible, still remains a challenge. 

\section{`Macroscopic' equation for the dragon-king non-linear dynamics}\label{section:MeSZG}

To describe the left-hand side of the $\lambda $-peak shown in Figs. \ref{figure:oba_zbocza} and 
\ref{figure:oba_zbocza_week}, we focus on the dynamics of the SZG node degree, $k_{SZG}$, in the frame of the continuum approach. In fact, the question is how its deterministic component (i.e. deprived fluctuations) or the first moment, $\bar{k}_{SZG}$, increases in time, if we assume (based on the empirical observation) that it monotonically increases. This means that, at least, a single edge is effectively attached to the SZG node at every time step. This edge comes from reservoir of remaining edges, which are not the members of the first coordination zone or layer of the SZG node. 

Continually motivated by the empirical data presented in Figs. \ref{figure:oba_zbocza} and \ref{figure:oba_zbocza_week} (by the erratic solid curve), our approach is divided into two stages. Within the first stage we deal with time ranging from $t_{crit}=164$[td]$\equiv $ 2005-08-11 to the middle of October 2006 (just before the last but one $k_{SZG}$ jump shown in Figs. \ref{figure:oba_zbocza} and \ref{figure:oba_zbocza_week}\footnote{The possible inflection point can be considered as a beginning of region of the impetuous $k_{SZG}$ increase. However, to find its site, the lower dispersion of empirical data is required. As suggested by our present empirical data (shown in Figs. \ref{figure:oba_zbocza} and \ref{figure:oba_zbocza_week}), we can only estimate that it is placed somewhere inside the period from 2006-08-01 (Tuesday) to 2006-10-03 (Monday). We denoted this possible day by the green dashed vertical line}). Within the second stage we consider the time range extending from the latter date to $t_{\lambda }=544$[td]$\equiv $ 2007-01-25 (the site of 
$t_{\lambda }$ is denoted by the red vertical dotted-dashed line). 

As we are looking for the dynamics of the first moment, $\bar{k}_{SZG}$, we disregarded fluctuations (similarly as it was made in Sec. VIII.1 in Ref. 
\cite{NGvKamp}) that is, we are looking for equation which relates only to the deterministic part of the corresponding Langevin equation. Hence, the generic coarse-grain or `macroscopic' equation of the system evolution, formally valid for both sub-periods, can be written in the clearly interpreted binomial form,
\begin{eqnarray}
\frac{\partial \bar{k}_{SZG}(t^{\prime})}{\partial t^{\prime }}&=&\sum_{l=1}^{n-1-\bar{k}_{SZG}(t^{\prime })} l\; p(l|\bar{k}_{SZG}(t^{\prime })) \nonumber \\
&=&\left(n-1-\bar{k}_{SZG}(t^{\prime })\right)b(1|\bar{k}_{SZG}(t^{\prime })),
\label{rown:genericdeq}
\end{eqnarray}
where  
\begin{eqnarray}
p(l|\bar{k}_{SZG}(t^{\prime }))&=&
\left(
\begin{array}{c}
n-1-\bar{k}_{SZG}(t^{\prime }) \\
l
\end{array}
\right) \nonumber \\
&\times &b(1|\bar{k}_{SZG}(t^{\prime }))^l\; (1-b(1|\bar{k}_{SZG}(t^{\prime }))^{n-1-\bar{k}_{SZG}(t^{\prime })-l}
\label{rown:trprob}
\end{eqnarray}
and
\begin{eqnarray}
t^{\prime }=
\left\{
\begin{array}{cc}
t-t_{crit}\; (\ge 0) & \mbox{for the first stage} \\ 
t-t_{\lambda }\; (<0 ) & \mbox{for the second stage}.
\label{rown:tcritlamb}
\end{array}
\right. 
\end{eqnarray}
Here, $l$ is an effective number of edges attached to the SZG node per unit time hence, both probabilities $p(\ldots )$ and $b(\ldots )$ are also the effective ones and can be interpreted as the corresponding rates. The binomial form of Eq. (\ref{rown:tcritlamb}) means that the edges are attached mutually independent - the only dependence is of the basic conditional probability per unit time, $b(1|\bar{k}_{SZG}(t^{\prime }))$  and 
$\bar{k}_{SZG}(t^{\prime })$.

The goal, herein, is to derive the dependence of the basic conditional probability $b(1|\bar{k}_{SZG}(t^{\prime }))$ on $\bar{k}_{SZG}(t^{\prime })$. Notably, this basic conditional probability was already studied in Sec. \ref{section:dbcond} but for much smaller vertex degree. We can expect that this conditional probability depends both on $\bar{k}_{SZG}(t^{\prime })$ and $n-1-\bar{k}_{SZG}(t^{\prime })$ numbers of edges. It is because, this probability can be considered as a quotient of two other probabilities. The first joint one, $b(1,\bar{k}_{SZG}(t^{\prime }))$, inversely proportional to 
$n-1-\bar{k}_{SZG}(t^{\prime })$, describes an event when single edge is randomly drawn from the reservoir. By the term `joint' we call the case where reservoir consists of $n-1-\bar{k}_{SZG}(t^{\prime })$ edges and simultaneously the star-like SZG superhub with $\bar{k}_{SZG}(t^{\prime })$ ones. The second probability, $p(\bar{k}_{SZG}(t^{\prime }))$, defines an event when $\bar{k}_{SZG}(t^{\prime })$ edges belong to the star-like SZG superhub. Indeed, this latter probability has to be separately considered for different phases. As we will see, this leads to the non-conserved order-parameter \emph{model C} dynamics (in the Hohenberg-Halperin terminology \cite{HH,HP}, which can be generally used), as the total number of vertices and edges is conserved for the entire MST network.

\subsection{Critical dynamics -- the first stage}\label{section:critdyn}

For the first stage we assume the second probability in the scaling form, that is, simply proportional to 
$[\bar{k}_{SZG}(t^{\prime })-\bar{k}_{SZG}(0)]^{\gamma }, \; \gamma > 0$, i.e. 
$p(\bar{k}_{SZG}(t^{\prime }))=\frac{1+\gamma }{A^{1+\gamma }}[\bar{k}_{SZG}(t^{\prime })-\bar{k}_{SZG}(0)]^{\gamma }$, which seems to be quite a natural choice\footnote{The simplest but marginal seems to be the case of exponent $\gamma =1$.} defining a critical dynamics. By the term `critical dynamics' we identify the dynamics, which leads, indeed, to the solution in the scaling form, explicitly involving the critical values of control parameters, i.e. valid for the scaling region. Hence, Eq. (\ref{rown:genericdeq}) takes the form related to the (slightly modified, deterministic) Allen-Cahn equation \cite{HP} (with properly defined a time-dependent diffusion coefficient),
\begin{eqnarray}
\frac{\partial \bar{k}_{SZG}(t^{\prime})}{\partial t^{\prime }}&=&\frac{A^{1+\gamma }}{1+\gamma }\; \frac{1}{[\bar{k}_{SZG}(t^{\prime })-\bar{k}_{SZG}(0)]^{\gamma }} \nonumber \\
&=&\frac{D}{\bar{k}_{SZG}(t^{\prime })-\bar{k}_{SZG}(0)},\; A, \gamma >0
\label{rown:simpleq1}
\end{eqnarray}
which has solution
\begin{eqnarray}
\bar{k}_{SZG}(t^{\prime })-\bar{k}_{SZG}(0)=A\; t^{\prime 1/z},\; z=1+\gamma ,
\label{rown:simpleq2}
\end{eqnarray}
where normalization constant $A$ together with exponent $\gamma $ usefully parametrized the prefactor in Eq. (\ref{rown:simpleq1}). Diffusion coefficient, D, assumed here the Arrhenius form $D=D_0\exp\left(\beta \varepsilon _{SZG}(t')\right)$, where $D_0=\frac{A^{1+\gamma }}{1+\gamma }$, the inverted temperature, $\beta $, is equivalent to $\gamma -1$, and the energy barrier, $\varepsilon _{SZG}(t')$, has logarithmic form 
$\ln \left(\bar{k}_{SZG}(t^{\prime })-\bar{k}_{SZG}(0) \right)$. The inverted temperature and energy are related to the corresponding ones defined in Sec. \ref{section:dbcond}. 

Although the diffusion coefficient is explicitly present in Eq. (\ref{rown:simpleq1}), this equation contains only time- and not space-dependent quantities as the short-range order parameter, $\bar{k}_{SZG}(t^{\prime })-\bar{k}_{SZG}(0)$, does not depend on space variables.

Indeed, this solution given by Eq. (\ref{rown:simpleq2}) was fitted to empirical data (forming the `Nucleation' range in Figs. \ref{figure:oba_zbocza} and \ref{figure:oba_zbocza_week}) and shown there by the green ($z=3$) and blue ($z=2$) solid curves. Apparently, no more details concerning the MST network was needed to obtain so good agreement. It suggest the universal character of the dynamic exponent $z$. This, we believe to be a result of dynamical criticality reached by the MST network. 

More precisely, in our case we deal with growth of the mean `droplet' size (or the SZG node's degree) during the nucleation process with dynamic exponent 
$z$, which is slowly decreasing function of time from the Lifshitz-Slyozov value $z=3$ down to $z=2$. This suggests that we simultaneously observe some coupled competitive growth processes \cite{BGC,GuSch} -- presumably, the nucleation and condensation ones. Notably, the Lifshitz-Slyozov growth dynamics is a conserved one, while in our case we deal with the non-conserved dynamics for the (local) short-range order parameter of the superstar-like SZG superhub. Nevertheless, the growth exponent $z=3$, which we observed, can be a reminiscence of the conservation of the total number of vertices and edges of the entire MST.

The simplest Langevin equation associated with Eq. (\ref{rown:simpleq1}) is a quasi-linear one (in the van Kampen terminology \cite{NGvKamp}), where the Langevin term describes the Gaussian white noise. This {Langevin} equation is equivalent to the quasi-linear Fokker-Planck equation. However, since amplitude $A$ in Eq. (\ref{rown:simpleq1}) is positive, the stationary solution of this Fokker-Planck equation does not exist. That is, drift and diffusion currents present there cannot mutually balance and we deal with a continuous transition from equilibrium scale-invariant network to the nucleating non-equilibrium one (see Figs. \ref{figure:oba_zbocza} and \ref{figure:oba_zbocza_week} for the illustration). 

What unusual, is that the logarithmic increase of $\bar{k}_{SZG}(t^{\prime })$ within the `Condensation' time range requires a different form of the probability that only $\bar{k}_{SZG}(t^{\prime })$ edges belong to the star-like SZG superhub. This will substantially modify Eq. (\ref{rown:simpleq1}).

\subsection{Diverging dynamics of a dragon king -- the second stage}\label{section:dddk}

We propose for the 'Condensation' time range, the probability that $\bar{k}_{SZG}(t^{\prime })$ edges belong to the star-like SZG superhub as proportional to exponential, i.e. $p\left(\bar {k}_{SZG}(t'))\right)=\frac{\tau _L}{A_L}\exp\left(-\bar{k}_{SZG}(t^{\prime })/A_L\right)$, where it is convenient to keep the proportionality constant as the ratio $\frac{\tau _L}{A_L}$. Hence, in the continuum limit, Eq. (\ref{rown:genericdeq}) takes the form
\begin{eqnarray}
\frac{\partial \bar{k}_{SZG}(t^{\prime })}{\partial t^{\prime }}=\frac{A_L}{\tau _L}\exp\left(\bar{k}_{SZG}(t^{\prime })/A_L\right),
\label{rown:lambda}
\end{eqnarray} 
where amplitude $A_L$ and relaxation time $\tau _L$, and center $t_{\lambda }$ (implicitly present in above equation) are positive quantities, found from the fit to empirical data (see Tab. \ref{table:Atltaushort} for daily and Tab. XX for weekly horizons).
\begin{table}
\begin{center}
\caption{Parameters characterizing left- and right-hand sides of $k_{SZG}$ $\lambda $-peak calculated for trading day horizon}
\label{table:Atltaushortd}
\vspace*{0.2in}
\begin{tabular}{|c|c|c|c|}
\hline
Side (J)& $A_J$ & $t_{\lambda }$[td] & $\tau _J$[td] \\
\hline \hline
$L$ & $14$ & $544$ & $2500$ \\
\hline
$R$ & $22$ & $544$ & $480$ \\
\hline
\end{tabular}
\end{center}
\end{table}
\begin{table}
\begin{center}
\caption{Parameters characterizing left- and right-hand sides of $k_{SZG}$ $\lambda $-peak calculated for trading week horizon}
\label{table:Atltaushortw}
\vspace*{0.2in}
\begin{tabular}{|c|c|c|c|}
\hline
Side (J)& $A_J$ & $t_{\lambda }$[td] & $\tau _J$[td] \\
\hline \hline
$L$ & $14$ & $110$ & $500$ \\
\hline
$R$ & $22$ & $110$ & $96$ \\
\hline
\end{tabular}
\end{center}
\end{table}

From Eq. (\ref{rown:lambda}) results that the SZG degree all the more increases per unit time the greater is the SZG degree at a given time. This means that `the richer becomes richer' is, herein, a conditional rule that the SZG vertex will be a richer in the next time step at condition that currently it is also rich. However, the probability to be so rich  (at a given time step) decreases exponentially with increasing of the SZG degree. Hence, it is difficult to become a rich node however, if it happened so, then the rate of edges' connection exponentially increases according with increase of the SZG degree.

It is a straightforward procedure to find a solution of Eq. (\ref{rown:lambda}) -- it takes, for $t_{\lambda }-t<\tau _L$, the following logarithmic form,
\begin{eqnarray}
\bar{k}_{SZG}(t^{\prime })=-A_L\; \ln \left(\frac{t_{\lambda }-t}{\tau _L}\right).
\label{rown:sollambda}
\end{eqnarray}
Apparently, this solution logarithmically diverges at the centre of $\lambda $-peak that is, at $t\rightarrow t_{\lambda }^-$. Indeed, this solution is well fitted to the empirical data (given by the erratic solid curve) and presented in Figs. \ref{figure:oba_zbocza} and \ref{figure:oba_zbocza_week} by the red solid curve. Hence, the postulate introduced at the beginning of this paragraphis valid for $\bar{k}_{SZG}\ge 40$, i.e. when superstar-like superhub is well formed.  Obviously, the result of Eq. (\ref{rown:sollambda}) type has a critical although marginal character in the sense, that it represents the case of vanishing critical dynamic exponent \cite{TKS}.

\subsection{Right-hand side of $\lambda $-peak -- the third stage}\label{section:rhlp}
 
The right-hand side of the $\lambda $-peak Eq. (\ref{rown:genericdeq}) should be confirmed by assuming (in agreement with empirical data) that the deterministic part of this side monotonically decreases with time. Hence, we correspondingly modify Eq. (\ref{rown:genericdeq}) as follows,
\begin{eqnarray}
\frac{\partial \bar{k}_{SZG}(t^{\prime})}{\partial t^{\prime }}=-\sum_{l=1}^{\bar{k}_{SZG}(t^{\prime })} l\; p(-l|\bar{k}_{SZG}(t^{\prime }))=-\bar{k}_{SZG}(t^{\prime })b(-1|\bar{k}_{SZG}(t^{\prime })), 
\label{rown:genericdeqR}
\end{eqnarray}
where we also express the probability per unit time, $p(-l|\bar{k}_{SZG}(t^{\prime }))$, of disconnection of $l$ edges from the SZG superhub at time $t^{\prime }=t-t_{\lambda }\; (>0)$, by the proper binomial representation, 
\begin{eqnarray}
p(-l|\bar{k}_{SZG}(t^{\prime }))&=&
\left(
\begin{array}{c}
\bar{k}_{SZG}(t^{\prime }) \\
l
\end{array}
\right) \nonumber \\
&\times &b(-1|\bar{k}_{SZG}(t^{\prime }))^l\; (1-b(-1|\bar{k}_{SZG}(t^{\prime }))^{\bar{k}_{SZG}(t^{\prime })-l}. 
\label{rown:trprobR}
\end{eqnarray}
Notably, the summation in Eq. (\ref{rown:genericdeqR}) is extended up to $\bar{k}_{SZG}(t^{\prime })$ edges, which simply means that all edges of the SZG node change (in average) their location in the network.

Analogously, as for the left-hand side of $\lambda $-peak, to solve Eq. (\ref{rown:genericdeqR}), the explicit dependence of the basic conditional probability per unit time, $b(-1|\bar{k}_{SZG}(t^{\prime }))$, on $\bar{k}_{SZG}(t^{\prime})$ is required. Again, this probability can be considered as a quotient of the other two probabilities. The first joint one per unit time, inversely proportional to $\bar{k}_{SZG}(t^{\prime })$, describes an event that given single edge is randomly drawn from the SZG superhub. The second probability defines an event that the star-like SZG superhub only consists of 
$\bar{k}_{SZG}(t^{\prime })$ edges. This latter probability was already proposed in the form of exponential in the previous paragraph. Here, we use it with conformed $A_R$ parameter that is, proportional to $\exp\left(-\bar{k}_{SZG}(t^{\prime })/A_R\right)$. 

Finally, in the continuum limit approach, Eq. (\ref{rown:genericdeqR}) takes the form
\begin{eqnarray}
\frac{\partial \bar{k}_{SZG}(t^{\prime })}{\partial t^{\prime }}=-\frac{A_R}{\tau _R}\exp\left(\bar{k}_{SZG}(t^{\prime })/A_R\right),
\label{rown:lambdaR}
\end{eqnarray} 
where amplitude $A_R$ and relaxation time $\tau _R$ are positive quantities, found from the empirical data (see Tabs. \ref{table:Atltaushortd} and 
\ref{table:Atltaushortw} for details).

The interpretation of above equation is analogous to Eq. (\ref{rown:lambda}) although, herein, we consider (due to the minus sign in its rhs) the rate of disconnection of edges from the superhub SZG. 

It is a straightforward procedure to find a solution of Eq. (\ref{rown:lambdaR}) -- it takes for 
$t-t_{\lambda }<\tau _R$ the following logarithmic form
\begin{eqnarray}
\bar{k}_{SZG}(t^{\prime })=-A_R\; \ln \left(\frac{t-t_{\lambda }}{\tau _R}\right).
\label{rown:sollambdaR}
\end{eqnarray}
Apparently, this solution also logarithmically diverges at the centre of $\lambda $-peak that is, at 
$t\rightarrow t_{\lambda }^+$. Indeed, this solution was fitted to the empirical data (given by the erratic solid curve) and presented in Figs. 
\ref{figure:oba_zbocza} and \ref{figure:oba_zbocza_week} by the solid blue curve.

Remarkably, the center $t_{\lambda }$ of the $\lambda $-peak obtained from (above mentioned) fits is common for both sides of the peak -- see Tabs. 
\ref{table:Atltaushortd} and \ref{table:Atltaushortw} (and also \ref{table:Atltau} and \ref{table:Atltauw} given in the next paragraph), which confirms self-consistencr of our approach. Remaining parameters were obtained as different ones although corresponding amplitudes have close values.

Our considerations provide the conclusion that complexity, present in the second equality in Eq. (\ref{rown:genericdeqR}), was reduced to the nonlinearity present in the corresponding Eqs. (\ref{rown:simpleq1}), (\ref{rown:lambda}), and (\ref{rown:lambdaR}).

A better empirical view at the $\lambda $-peak, we gain in Sec. \ref{section:aopar} by applying complementary, relative, time-dependent short-range order parameter that is, the difference $\Delta _{SZG}(t)=k_{SZG}(t)-k_2(t)$, where $k_2(t)$ is degree of a temporal vice-leader.

\subsection{$\lambda $-peak and condensation within complementary order parameter}\label{section:aopar}

In this Section we consider, by using above defined complementary short-range order parameter $\Delta _{SZG}(t)$, the dynamics of the richest SZG vertex -- or dragon king -- in the vicinity of January 25, 2007\footnote{Degree of the SZG vertex reaches its maximal value equals $k_{SZG}^{max}=88$ at January 25, 2007.} as its evolution there is much more distinct and intriguing than for other rich vertices. 

\begin{figure}
\begin{center}
\bigskip
\includegraphics[width=140mm,angle=0,clip]{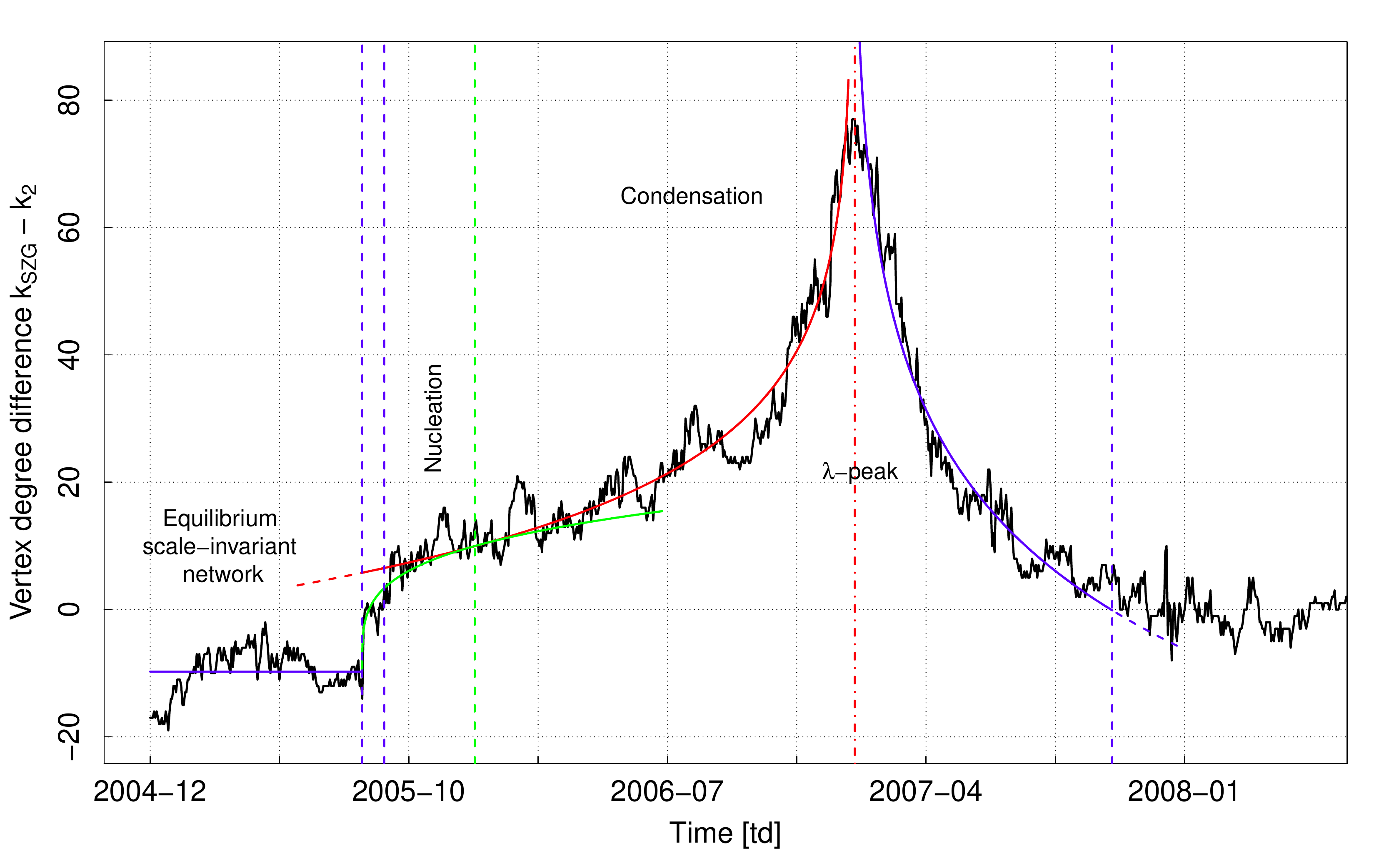}
\caption{The empirical, temporal vertex degree difference $k_{SZG}(t)-k_2(t)$ (black erratic solid curve obtained for trading day horizon) vs. time $t$, which forms the $\lambda $-peak -- time plays the role of the control parameter. The smooth well fitted red and blue diverging solid curves 
were obtained by using the function $-A_J\ln \left(|t-t_{\lambda }|/\tau _J\right),\; J=L, R$, for properly chosen values of parameters $A_J$ and 
$\tau _J$ for lhs ($J=L$)- and rhs ($J=R$)-hand sides of $\lambda $-peak (cf. Table \ref{table:Atltau}). The critical (transition) time, 
$t_{\lambda }$, is common for both sides, as it is required for self-consistency by the interpretation of $\lambda $-peak. Central vertical red dashed-dotted line denotes the site of $t_{\lambda }=t_{min}^{MOL}=544$[td]$\equiv $2007-01-25 (Thursday). The above mentioned diverging curves illustrate, what we call, the dynamic $\lambda $-transition. Additionally, both green solid curves illustrate the nucleation process defined by the Expression 
(\ref{rown:simpleq2}). For the early stage critical dynamics, the dynamic exponents were restricted by boundary values $z=3.0$ and $4.0$ (with amplitude $A$ equals $4,80$ and $6.45$, respectively). It seems that the true exponent is placed somewhere in between these two -- unfortunately, for the current empirical data, we are not able to achieve sufficient accuracy to confirm this. One of these curves (for $z=4$ and $A=6.45$) was paired with the red one at intersection defined by the green dashed vertical line located at 2005-12-12 (Friday; $255$[td]). The horizontal blue straight line has a height equal to $-9.73$ that is, an average value of all daily time-series empirical data for the period from 2004-12-01 (Wednesday) to 2005-08-12 (Friday).}
\label{figure:kondensat_day}
\end{center}
\end{figure}

In Figs. \ref{figure:kondensat_day} and \ref{figure:kondensat} the short-range order parameter, $\Delta _{SZG}(t)$, is plotted\footnote{The maximal value of degree difference equals $\Delta_{SZG}^{max}=77$ and is located at January 25, 2007.} for daily and weekly horizons, respectively. The second degree difference, $\Delta _{SZG}^2(t)=k_2(t)-k_3(t)$, (where $k_3(t)$ is degree of the company temporaly occupying the third position in the rank at time $t$) almost vanishes within the range of the peak (cf. Fig. 6 in Ref. \cite{WSGKS}). The predominant role of SZG company as a dragon-king is, therefore, evident. It should be emphasized that this dynamic peak is also of $\lambda $ type, as both of its sides are well fitted by the function 
$-A\ln \left(|t-t_{\lambda }|/\tau \right)$, where $|t-t_{\lambda }| < \tau $. The values of parameters $A, t_{\lambda }$ and $\tau $ for the left- and right-hand sides of the peak for daily and weekly horizons are shown in Tabs. \ref{table:Atltau} and \ref{table:Atltauw}, respectively. Apparently, the critical (transition) time (or threshold) $t_{\lambda }=t_{min}^{MOL}=544$[td]$\equiv 2007-01-25$, where $t_{min}^{MOL}$ is the time when MOL becomes minimal. Fortunately, the existence of this dynamical $\lambda $-peak confirms our earlier observations, e.g., concerning the most significant one, i.e. the peak location  (cf. Sec. \ref{section:critdyn} -- \ref{section:rhlp}) or the common centre of the peak.
 
\begin{figure}
\begin{center}
\bigskip
\includegraphics[width=150mm,angle=0,clip]{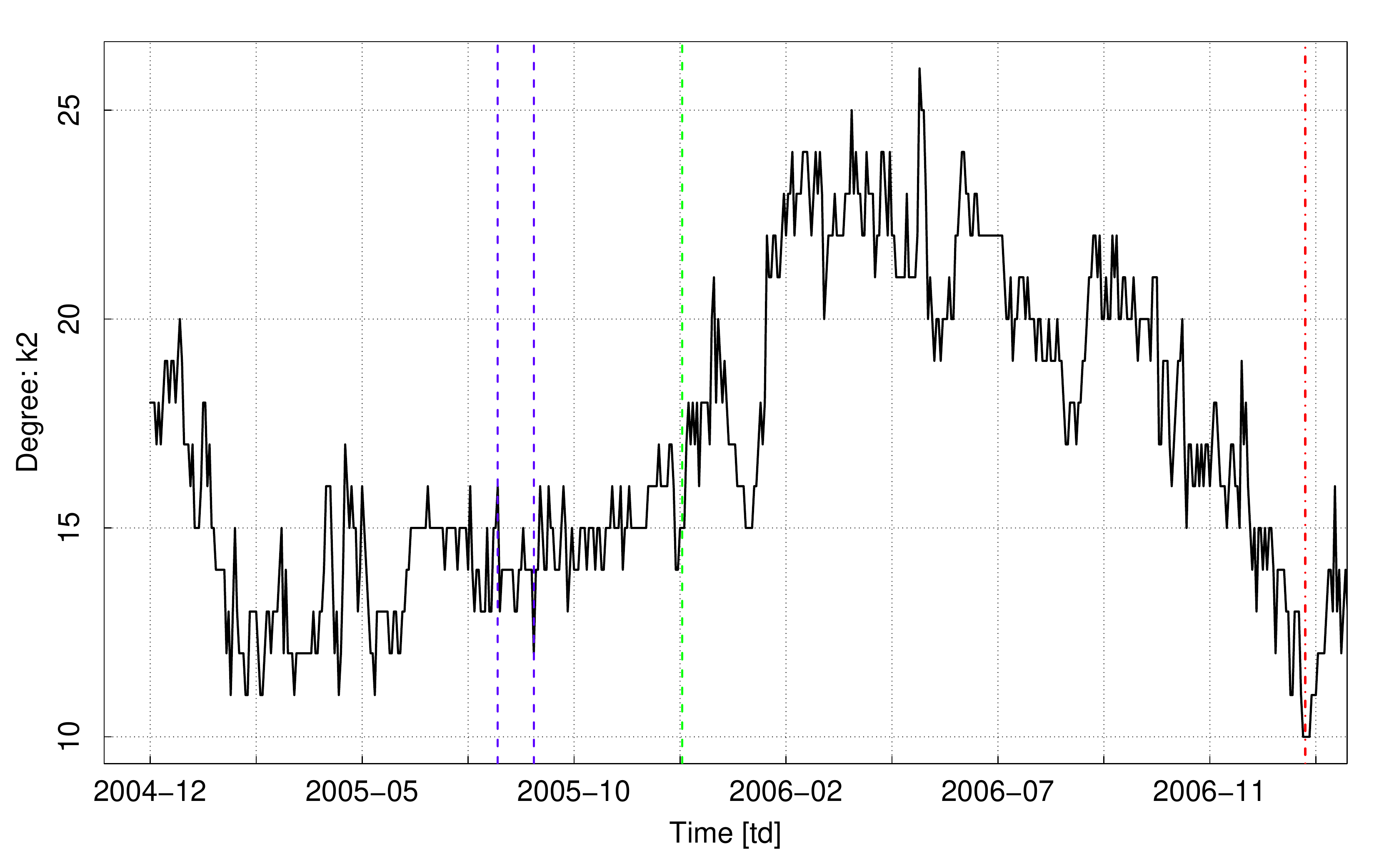}
\caption{The empirical temporal vice-leader dependence on time (an erratic solid curve obtained for trading day horizon). The  
SWV vertex degree, $k_{SWV}(t)$,  vs. time $t$ (counted in trading days) concerns the last segment of the curve.}
\label{figure:k2_left}
\end{center}
\end{figure}

\begin{figure}
\begin{center}
\bigskip
\includegraphics[width=140mm,angle=0,clip]{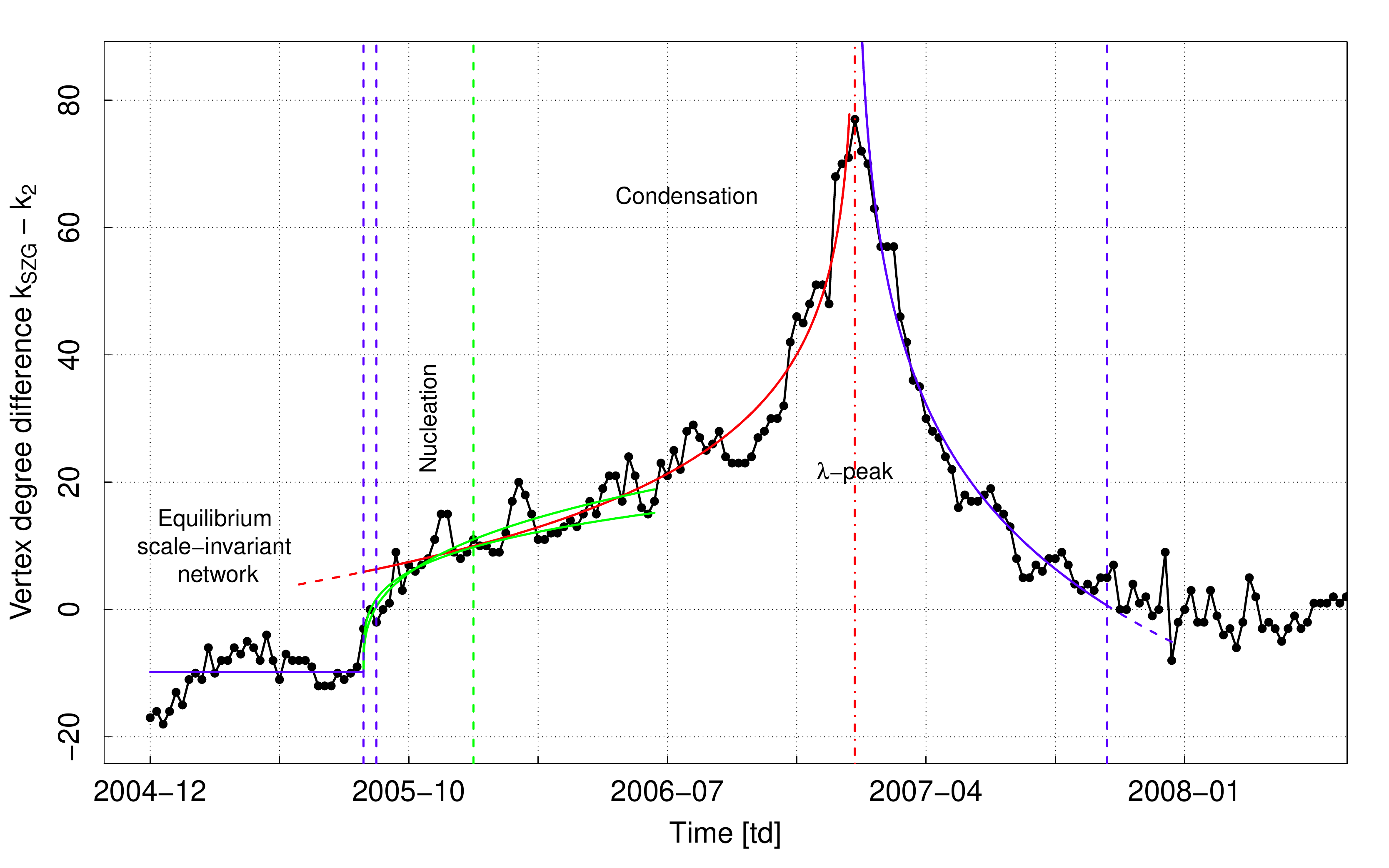}
\caption{The empirical, temporal vertex degree difference $k_{SZG}(t)-k_2(t)$ (small black circles connected by erratic solid curve obtained for trading week horizon) vs. time $t$ (plotted in trading days), which forms a dynamic $\lambda $-peak -- time plays the role of the control parameter. The solid fitted curves (red and blue with short dashed bonds) were obtained by using the function $-A_J\ln \left(|t-t_{\lambda }|/\tau _J\right)$ with properly chosen values of parameters $A_J$ and $\tau _J$ for the left- and right-hand sides of the peak, i.e. for $J=L$ and $J=R$, respectively (cf. Table 
\ref{table:Atltauw}). The transition (critical) time, $t_{\lambda }$, is common for both sides of the peak, as is required. The red vertical dashed-dotted line denotes the $t_{\lambda }=t_{min}^{MOL}=550/5=110$[tw]$\equiv $ 2007-01-29 (Monday). The plot illustrates, what we call, a dynamic 
$\lambda $-transition at $t_{\lambda }$. Additionally, both green curves illustrate the nucleation process defined by Expression 
$A_0+A(t-t_{crit}^{1/z}$, for $t\geq t_{crit}=170/5=34$[tw]$\equiv $ 2005-08-15 (Monday), analogously to Eq. 
(\ref{rown:simpleq2}). For the early stage of critical dynamics, the dynamic exponent is limited by its boundary values $z=3.0$ and $4.0$ (with amplitude $A$ equals $4,80$ and $6.45$, respectively). It seems that true dynamic exponent is placed somewhere in between these two -- however, to distinguish it a  better accuracy is required, which cannot be achieved for the current empirical data. One of these curves (for $z=4$ and $A=6.45$) was paired with the red one at the intersection defined by the green dashed vertical line located at 2005-12-12 (Friday; $51$[tw]) with green and red curves. The horizontal blue straight line has height equals $-9.79$ that is, an average value of all weekly time-series empirical data for the period from 2004-12-01 (Wednesday) to 2005-08-15 (Monday). Indeed, the latter date defines 
$t_{crit}=34$[tw], i.e. the site of the first blue vertical dashed line. The place of the second blue vertical dashed line (defined in Sec. \ref{section:Dpaed}) is given by 2005-08-29 (Monday; $36$[tw]). The blue vertical dashed line at the right-hand side of $\lambda $-peak, placed at 2007-10-29 (Monday; $149$[tw]), is defined in Sec. \ref{section:Dpaed}.}
\label{figure:kondensat}
\end{center}
\end{figure}

\begin{table}
\begin{center}
\caption{Parameters characterizing left- and right-hand sides of the $\Delta _{SZG}$ $\lambda $-peak calculated for trading day horizon}
\label{table:Atltau}
\vspace*{0.2in}
\begin{tabular}{|c|c|c|c|}
\hline
Side (J)& $A_J$ & $t_{\lambda }$[td] & $\tau _J$[td] \\
\hline \hline
$L$ & $16$ & $544$ & $544$ \\
\hline
$R$ & $25$ & $544$ & $200$ \\
\hline
\end{tabular}
\end{center}
\end{table}

\begin{table}
\begin{center}
\caption{Parameters characterizing left- and right-hand sides of the $\Delta _{SZG}$ $\lambda $-peak calculated for trading week horizon}
\label{table:Atltauw}
\vspace*{0.2in}
\begin{tabular}{|c|c|c|c|}
\hline
Side (J) & $A_J$ & $t_{\lambda }$[tw] & $\tau _J$[tw] \\
\hline \hline
$L$ & $16$ & $110$ & $110$ \\
\hline
$R$ & $25$ & $110$ & $40$ \\
\hline
\end{tabular}
\end{center}
\end{table}

Apparently, the temporal short-range order parameter $\Delta _{SZG}(t)$ is better suited to study the left-hand side of 
$\lambda $-peak than $k_{SZG}(t)$. However, the latter one makes possible a more refined study of the nucleation process. Hence, there are two complementary views on the same phenomena, which makes the analysis more versatile.

\section{Discussion and concluding remarks}\label{section:results}\label{section:Sumconcl}

In spite of the perceived importance of the dynamical phase transitions on the financial markets, in particular as applied to the analysis of market crashes, a systematic empirical and phenomenological analysis of this phenomenon is still incomplete, with the reality invalidating the established views and contradicting the established facts.

By using the canonical MST network we studied, as a representative example, the dynamics of the Frankfurt Stock Exchange (as a socio-economical thermometer of a leading (German) economy in Europe), in an attempt to fill this void. Other European stock exchanges (e.g., the Warsaw Stock Exchange) are following FSE, particularly during  the recent worldwide financial and economical crisis, still enduring. The amazing similarity of many stock exchange indices has been earlier convincingly visualized and briefly discussed by Didier Sornette (cf. Fig. 24 in Ref. \cite{DSDS}). 

The most significant results of our work were provided in Figs. \ref{figure:oba_zbocza} and \ref{figure:kondensat_day}, where several MST network states, forming the dynamical, structural, and topological phase transitions are clearly shown. In these figures we completed the dynamical phase diagrams, showing three kinds of the continuous phase transitions, summarized below and accompanied by the related highlights. 

The initial phase transition occurs from the equilibrium to the non-equilibrium MST networks at some critical time 
$t_{crit}(\equiv $2005-08-11 (Thursday)). Then the coalescence of edges with the FSE temporal leader SALZGITTER (SZG) AG-Stahl und Technologie company is observed within the nucleation characterized, in approximation, by the Lifsthiz-Slyozov growth exponent (although we deal, herein, with the non-conserved dynamics). In the meantime\footnote{The transition region (between nucleation and condensation ones) is shown (to good approximation) in Fig. 
\ref{figure:SZG_k2_correct} by the longest horizontal curve.}, before the third phase transition, the nucleation accelerates and transforms to the condensation process, finally forming a logarithmically diverging $\lambda $-peak of $k_{SZG}$ and $k_{SZG}-k_2$ at the subsequent critical time ($t_{\lambda }=544$[td]$\equiv $2007-01-25 (Thursday)). Next, over three quarters, the peak logarithmically decreases (up to some reasonably small fluctuations) resulting in decentralized graphs -- see Fig. \ref{figure:ZEP_558} for details, where disintegration of the super-star like graph (or superhub) to several loosely connected clusters (or subgraphs) having their owns local leaders, are clearly seen.

A more detailed explanation of above brief considerations is also provided. The reference MST network states -- denoted in Figs. \ref{figure:oba_zbocza} and \ref{figure:kondensat_day} by the term `Equilibrium scale-invariant network' -- are located in the first sub-period from 2004-12-05 (Monday) to 2005-08-11 (Thursday). Their typical, hierarchical scale-invariant structure is presented in a comprehensive form in Fig. \ref{figure:ZE_16}. It is characterized by the power law distribution of its vertex degrees (driven by exponent $\alpha =2.98$). Apparently, there is no vertex located off the power law (except for the first less important boundary case as a result of a finite size effect). Indeed, the MST network equilibrium states (or the equilibrium phase) are defined by
\begin{itemize}
\item[(i)] detailed balance conditions (\ref{rown:currentout0}) and (\ref{rown:currentin0}), hence, 
\item[(ii)] by the properly suited transition probabilities given by Eqs. (\ref{rown:trdegreed}) and (\ref{rown:finaltrans}), together with, 
\item[(iii)] the sojourn probabilities (\ref{rown:normhk}), where time-independent equilibrium power law distributions of vertex degrees used, are governed by the average exponent $\bar{\alpha }=3.07$ (for details see Fig. 
\ref{figure:mean_alpha}). 
\end{itemize}
The agreement of these transition and sojourn probabilities with their empirical counterparts is well established (cf. Fig. \ref{figure:deg_ch_prob-1} for details). Furthermore, the results plotted in Fig. 
\ref{figure:deg_ch_prob-1} are generally valid  for the `fluid' of vertices having degrees not exceeding a dozen or so, and not only for the MST network in equilibrium. In other words, they are valid for any sub-periods, since the remaining non-equilibrium `fluid' is well separated, as it consists of at most few vertices having much larger degrees, coming off the equilibrium power law (see, e.g., Fig. \ref{figure:ZEP_341} for details). Hence, the two-fluid modeling of the MST network evolution is supported, in a crude analogy to the II $ ^4$He superfluidity.  

The considerations of subsequent sub-periods were based on the amazing observation that the peripheral SZG company became a dominant one, barely within two trading days 2005-08-11/12 (Thursday/Friday) and 2005-08-12/15 (Friday/Monday), which results in the critical-like change of the MST structure (see Figs. \ref{figure:ZE_22}, \ref{figure:ZE_23}, and \ref{figure:ZEP_24} for comparison as well as Figs. \ref{figure:oba_zbocza} and 
\ref{figure:kondensat_day} for the corresponding phase diagrams). This change gives the characteristic dependence of the SZG vertex degree, $k_{SZG}$, and the vertex degree difference, $k_{SZG}-k_2$, vs. time as shown in Figs. 
\ref{figure:oba_zbocza}, \ref{figure:oba_zbocza_week}, \ref{figure:kondensat_day}, and \ref{figure:kondensat}, where the first critical point at $t_{crit}=164$[td]$\equiv $2005-08-11 (denoted by the blue dashed vertical line) is well seen. Notably, the dependence obtained from our deterministic coarse-grain ('macroscopic') Eq. (\ref{rown:genericdeq}) in the form
\begin{eqnarray}
\left\{
\begin{array}{cc}
A_0, & \mbox{for $t<t_{crit}$} \nonumber \\
A_0+A(t-t_{crit})^{1/z},\; z>1, & \mbox{for $t\geq t_{crit},$} \nonumber \\
\end{array}
\label{rown:A0zttc}
\right.
\end{eqnarray}
valid both for $k_{SZG}$ and $k_{SZG}-k_2$ vs. $t$, is a continuous one even at $t=t_{crit}$, although its derivative diverges at the critical point. Therefore, we can say that this is a reminiscent of the dynamical structural continuous (at least of the 2nd order) phase transition. This is a transition between the scale-invariant network in equilibrium present for $t < t_{crit}$ 
(as discussed above) and the non-equilibrium phase of the complex network present for $t_{crit}\leq t\leq t_{nucl}=286$[td]$\equiv $2006-01-30, where 
$t_{nucl}$ is denoted in Figs. \ref{figure:oba_zbocza}, \ref{figure:oba_zbocza_week}, \ref{figure:kondensat_day}, and \ref{figure:kondensat} by the green vertical dashed lines. 

The non-equilibrium phase is characterized by the nucleation process of edges coalesced by the dominant SZG superhub. This process (a slow mode or slow growth law) can be considered for $z=3$ (the solid green curves\footnote{In Fig. \ref{figure:kondensat_day}, the green curve, corresponding to exponent $z=3$, runs initially slightly above the second green curve concerning $z=4$, but finally it runs below.} in both figures) as related to the Lifshitz-Slyozov canonical process of a droplet growth \cite{BGC} as characterized by the same dynamical (growth law) exponent although, we deal here with a non-conserved dynamics, but at a fixed total number of vertices and edges. This nucleation could also serve as an analogy to the non-equilibrium ordering kinetics 
\cite{TGLW} despite the complexity of a growth process and non-linear driving forces involved. The coalescence process for our case is slow (slow mode) because the richest vertices are located still relatively far (counted in `handshakes') from the SZG node (cf. Figs. \ref{figure:SZG_k2_correct} and 
\ref{figure:SZG_k3_correct}). Therefore, the coordination zones of the SZG node have insufficient number of vertices to accelerate an attachment process. It is worthwhile to relate this transition to nucleation, with some significant destructive events on worldwide markets. Our conjecture is that, the at least crash on the market of new US houses for sale in July 2005 together with a giant fluctuation on US car market at that time, had become a crash catalyzer a month later.

The slow mode transforms to a fast mode for $t_{nucl} < t < t_{\lambda }$, where location of $t_{\lambda }$ is denoted by the red vertical dashed-dotted line. The slow and fast modes are shown by the green and red curves, respectively, pieced together before the `Condensation' region exhibited in Figs. 
\ref{figure:oba_zbocza}, \ref{figure:oba_zbocza_week}, \ref{figure:kondensat_day}, and \ref{figure:kondensat}. The red solid curves shown in the figures were obtained by fitting the function $-A_L\ln \left(\left(t_{\lambda }-t\right)/\tau _L\right),\; \tau _L=t_{\lambda }$, as a solution of Eq. 
(\ref{rown:lambda}). For the `Condensation' region we observed that the richest vertices are effectively `attracted' by the SZG superhub making its first, second and the third coordination zones richer and richer in edges during the MST evolution. Indeed, this mechanism induces impetuous, logarithmic coalescence of edges to the SZG superhub (or temporary attractor) leading to a condensate. Obviously, all these phenomena somehow relate to the bad news from the US economy (and not only from the selected markets) accumulated throughout the year 2006.

However, for time $t > t_{\lambda }$ the condensate is logarithmically decaying, in accordance with the function 
$-A_R\ln \left(\left(t-t_{\lambda }\right)/\tau _R\right),\; \tau _R\neq t_{\lambda }$), as a solution of Eq. (\ref{rown:lambdaR}), 
and disintegrates for the few loosely connected clusters, or sectors, centralized around their leaders. That is, for the later time, the SZG superhub is logarithmically decreasing, losing its edges (see Fig. \ref{figure:ZEP_558}) and finally seizes to be the dragon-king. This makes the final MST network phase step-by-step more similar to the initial `Equilibrium scale-invariant network', closing a single (quasi-)cycle of the evolution of a market. 

The non-equilibrium evolution of the entire complex network is reduced, herein, to the non-linear dynamics of the dragon-king or super-extreme event, which makes its solution the effective one. Noticably, we are saying about its deterministic component, leaving its stochastic part to the subsequent work, where complete Langevin equation is discussed in the context.

We can conclude that we have demonstrated the first evidence for real-life condensation phenomenon with an associated $\lambda $-transition, together with a preceding phase of nucleation growth. Further, we have investigated in empirical and phenomenological ways, a complete phase diagrams for these intriguing dynamic phase transitions in real-world complex networks. We expect that our results will be inspiring for interdisciplinary physicists involved in a broad spectrum of disciplines studying emergence evoked by complexity. Furthermore, we expect that our work, providing a sophisticated example of the dynamical paradigm of phase transitions and critical phenomenon (concerning a network as complex as stock market), will provide a new impulse to develop a unicersal theory of phase transitions in evolving complex networks.


\end{document}